\documentclass{elsart}
\usepackage{epsfig}
\usepackage{amsmath}
\usepackage{color}
\usepackage{amssymb}
\usepackage{graphicx}
\usepackage{subfigure}
\usepackage{rotating}
\usepackage{setspace}

\newcommand \be{\begin{equation}}
\newcommand \ba{\begin{eqnarray}}
\newcommand \ee{\end{equation}}
\newcommand \ea{\end{eqnarray}}

\begin{document}


\begin{frontmatter}
\title{Testing the Stability of the 2000 US Stock Market ``Antibubble''}
\author[ecust,igpp]{\small{Wei-Xing Zhou}},
\author[igpp,ess,nice]{\small{Didier Sornette}\thanksref{EM}}
\address[ecust]{State Key Laboratory of Chemical Reaction
Engineering, East China University of Science and Technology,
Shanghai 200237, China}
\address[igpp]{Institute of Geophysics and Planetary Physics,
University of California, Los Angeles, CA 90095}
\address[ess]{Department of Earth and Space Sciences, University of
California, Los Angeles, CA 90095}
\address[nice]{Laboratoire de Physique de la Mati\`ere Condens\'ee,
CNRS UMR 6622 and Universit\'e de Nice-Sophia Antipolis, 06108 Nice
Cedex 2, France}
\thanks[EM]{Corresponding author. Department of Earth and Space
Sciences and Institute of Geophysics and Planetary Physics,
University of California, Los Angeles, CA 90095-1567, USA. Tel:
+1-310-825-2863; Fax: +1-310-206-3051. {\it E-mail address:}\/
sornette@moho.ess.ucla.edu (D. Sornette)\\
http://www.ess.ucla.edu/faculty/sornette/}

\begin{abstract}
Since August 2000, the stock market in the USA as well as most
other western markets have depreciated almost in synchrony
according to complex patterns of drops and local rebounds. In
\cite{SZ02QF}, we have proposed to describe this phenomenon using
the concept of a log-periodic power law (LPPL) antibubble,
characterizing behavioral herding between investors leading to a
competition between positive and negative feedbacks in the pricing
process. A monthly prediction for the future evolution of the US
S\&P 500 index has been issued, monitored and updated in
\cite{urlprediction}, which is still running. Here, we test the
possible existence of a regime switching in the US S\&P 500
antibubble. First, we find some evidence that the antibubble
has exhibited a transition in log-periodicity described by a
so-called second-order log-periodicity. Second, we develop a
battery of tests to detect a possible end of the antibubble of the
first order which suggest that the antibubble was alive in August
2003 but has ended in the USA, when expressed in the local US dollar
currency. Our tests provide quantitative measures to diagnose the
end of an antibubble. Such diagnostic is not instantaneous and
requires from three to six months within the new regime before
assessing its existence with confidence. From the perspective of
foreign investors in their currencies (S\&P500 denominated in
British pound or in euro) or when expressed in gold so as to
correct for an arguably artificial US\$ valuation associated with
the Federal Reserve interest rate and monetary policy, we find
that the S\&P 500 antibubble is still alive and running its
course. Similar analyses performed on the major European stock
markets (CAC 40 of France, DAX of Germany, and FTSE 100 of United
Kingdom) show that the antibubble is also present and continuing there.
\end{abstract}


\end{frontmatter}

\typeout{SET RUN AUTHOR to \@runauthor}

\section{Introduction}

In 1999, in order to describe the evolution of the Japanese stock
market since its all-time high in December 1989, Johansen and
Sornette introduced the concept of an ``antibubble'' as a
counterpart of a bubble resulting from the same herding behavior
and characterized by log-periodic power-law (LPPL) structures but
with decelerating (rather than accelerating) oscillations
\cite{JS99IJMPC}. The term ``antibubble'' is inspired by the
concept of ``antiparticle'' in physics. Just as an antiparticle is
identical to its sister particle except that it carries exactly
opposite charges and destroys its sister particle upon encounters,
an antibubble is both the same and the opposite of a bubble; it's
the same because similar herding patterns occur, but with a mostly
bearish versus bullish slant. Some antibubbles can also describe
increasing markets over long times, although a bearish phase is
more commonly recognized in the markets
\cite{ZS02Global,J03QF,SZ03QF,GM03}. In August 2002, we detected
the existence of a clear signature of an antibubble in the
relaxation of the US S\&P 500 index since August 2000 with high
statistical significance, in the form of strong log-periodic
components \cite{SZ02QF}. Similarly to the prediction offered in
\cite{JS99IJMPC} for the evolution of the Nikkei index which was
later evaluated in \cite{evaljohsor}, we presented a prediction
for the future evolution of the US S\&P 500 index
\cite{SZ02QF,SZ03QF}. This prediction has been monitored and
updated once a month at the URL \cite{urlprediction}. Accompanying the US
stock markets, the antibubble regime since 2000 seems to be a
world-wide phenomenon in the major western stock market
\cite{ZS02Global}. These works on antibubbles extend a large
amount of theoretical and empirical work on LPPL bubbles which
often end in crashes or strong corrections (see
\cite{Bookcrash,CMC,JS02XXX,sorzhoupat} and references therein).
In this context, Roehner has investigated the resilience pattern
around large price peaks \cite{R00EPJB} and has found strong
negative correlations between stock market crash-recovery and
interest rate spread \cite{R00IJMPC}.

In contrast to a LPPL bubble whose end is automatically described
by one of the parameters, the critical time $t_c$, the LPPL
formulation of an antibubble does not say anything a priori about
its duration. For prediction purpose, the agonizing question is
whether the detection of an antibubble pattern ensures its
continuation in the future and for how long. In the case of the
Japanese antibubble studied in depth \cite{JS99IJMPC,evaljohsor},
the detection was performed in early January 1999, corresponding
to 9 years since the birth of the antibubble. The prediction
issued in early January 1999 turned out to be followed
subsequently ex-ante by the Nikkei index over more than two years.
However, in January 1999, it was hard a priori to assess for how
long the theory would be a correct predictor of the future
evolution of the Nikkei index.

Here, we address this question of the detection of a change of
regime from an antibubble phase to something else. For this
purpose, the present situation is perhaps more favorable than for
the Nikkei in January 1999 for the following reason. As we said
above, in January 1999, the antibubble has been unfolding itself
already for 9 years. It was found necessary to extend the LPPL
theory from a first-order log-periodic formula to a second-order
and then to a third-order formula. It should be stressed that the
first-order formula is embedded as a special case of the
second-order formula which is itself embedded as a special case of
the third-order formula.  The logic of this succession of formulas
is that they represent successive improvement to describe the
market price at time intervals further and further away from the
early development of the antibubble. The larger is the order of
the formula, the larger is the time interval over which the theory
applies. The prediction issued in January 1999 was performed based
on the third-order LPPL formula. In contrast, the analysis of the
US S\&P 500 antibubble has been performed much earlier after about
only $2.5$ years since its inception in August 2000 \cite{SZ02QF}.
Due to this relatively short time span, it was found that the
first-order formula was sufficient to describe the empirical data,
while the second-order (and a fortiori the third-order) formula
was not needed as it did not lead to any statistically significant
improvement. We thus concluded that the S\&P 500 index had not yet
entered into the second phase in which the angular log-frequency
may start its shift to another value, as did the 1990 Nikkei
antibubble after about 2.5 years. However, this situation offers
the possibility for tracking a possible future change of regime
from the first-order to the second-order formula. This is the
first purpose of this paper. Using data garnered over ten additional
months, we show
that one can start
to detect the occurrence of such a change of regime. Adding an
additional year of
data confirms further this conclusion, as we shall show. The
statistical tests described below give the probability to reject
the hypothesis that the market has not entered the second phase in
which the angular log-frequency is shifting to another value.
These results suggest the possibility that, indeed, we have
entered a cross-over regime in log-frequency shift. The improved
second-order log-periodic formula has implications in the prediction
of the future drops of the markets.

The second purpose of our paper is to develop a battery of tests
to gain a better understanding of which scenario might be the most
likely to unfold: is the antibubble likely to continue and is the
market when expressed in one of the major foreign currencies or in gold
to drop further? Or will the
stock market transit to another regime, perhaps rebound to develop
a new bullish regime? Or even worse (from the point of view of our
model): is it possible that the US and European stock market has already
entered a regime different from that described by the LPPL
antibubble and that we have not yet taken this into account in our
updates presented in \cite{urlprediction}? The present paper provides
the theoretical basis and the statistical anchor underlying the
monthly prediction updates which are available at the URL \cite{urlprediction}.

\section{Angular log-periodic frequency shifting?}
\label{s1:Landau12}

\subsection{First- and second-order LPPL formulas}

Let us begin by recalling the
mathematical expression
of the price evolution trajectory of an antibubble \cite{JS99IJMPC,SZ02QF}:
\begin{equation}
\ln[p(t)] = A + B\tau^m + C\tau^m
\cos\left[\omega\ln(\tau)+\phi\right] + \epsilon(t)~,
\label{Eq:lnpt}
\end{equation}
where $p(t)$ is the price, $\epsilon(t)$ is the noise or fit
residuals, $\omega$ is the angular log-frequency, $B<0$ for
(bearish) antibubbles of interest here, $m$ is positive to ensure
a finite price at the critical initiation time $t_c$ of the
antibubble, $\phi$ is a phase which can be absorbed in a
re-definition of the unit of the time, and $\tau=t-t_c$ is the
distance to the critical time $t_c$ or onset of the antibubble.

As explained for instance in
\cite{Bookcrash,CMC,JS02XXX,sorzhoupat}, the power law
acceleration $B\tau^m$ and log-periodicity
$\cos\left[\omega\ln(\tau)-\phi\right]$ are both intimately linked
to behavioral herding of agents whose investments involve a
competition between positive and negative feedbacks
\cite{Idesor,sorIde} leading to a critical point. Close to
criticality, the ``order parameter'' $F \equiv \ln[p(t)]$ can be
expanded according to a so-called Landau expansion
\cite{SJ97PhysA} as a function of the ``control parameter'' $\tau$
\begin{equation}
     \frac{dF(\tau)}{d\ln \tau} = \alpha F(\tau)+\beta |F(\tau)|^2F(\tau)...~,
     \label{Eq:Landau0}
\end{equation}
where the coefficients $\alpha$ and $\beta$ can, on general
ground, be complex. Starting with the Landau expansion close to
the critical point $\tau=0$, keeping only the first-order term
$\alpha F(\tau)$ retrieves Eq.~(\ref{Eq:lnpt}). Inclusion of the
second-order term $\beta |F(\tau)|^2F(\tau)$ leads to
\cite{SJ97PhysA}
\begin{equation}
\ln[p(t)] \approx A + \frac{B\tau^m+ C\tau^m\cos\left\{\omega\ln
\tau + \frac{\Delta_\omega}{2m}\ln \left[1+\left(\frac{\tau}
{\Delta_t}\right)^{2m}\right] +\phi\right\}} {\sqrt{1+
\left(\frac{\tau} {\Delta_t}\right) ^{2m}}}~, \label{Eq:Landau2}
\end{equation}
where $\Delta_\omega \to 0$ and $\Delta_t \to \infty$ for $|\beta|
\to 0$. Higher-order terms in the Landau expansion
(\ref{Eq:Landau0}) can be taken into account to describe the
behavior at long times further away from the critical point $t_c$,
as done for the Nikkei antibubble \cite{JS99IJMPC}.

\subsection{Statistical tests}

The question, whether the S\&P 500 index has entered or not the
second phase in which the angular log-frequency is shifting to
another value, amounts to comparing the quality of the fits of the
data with the first-order (\ref{Eq:lnpt}) and with the
second-order formula (\ref{Eq:Landau2}).

Since the hypothesis that the S\&P 500 index follows the
first-order formula (\ref{Eq:lnpt}) is imbedded within the
hypothesis that it follows the second-order formula
(\ref{Eq:Landau2}), we can use the general theory of nested
hypothesis testing. Calling $\chi_1$ and $\chi_2$ the
root-mean-squares of OLS (ordinary least-squares) fits with
(\ref{Eq:lnpt}) and (\ref{Eq:Landau2}) of the price time series of
the S\&P 500 index with length $n$, the likelihood-ratio or Wilks
test states that the log-likelihood-ratio
$T(\kappa)=2n\ln(\chi_1/\chi_2)$ follows the chi-square
distribution with $\kappa=2$ degrees of freedom, asymptotically
when $n$ tends to infinity.

\begin{figure}
\centering
\includegraphics[width=12cm]{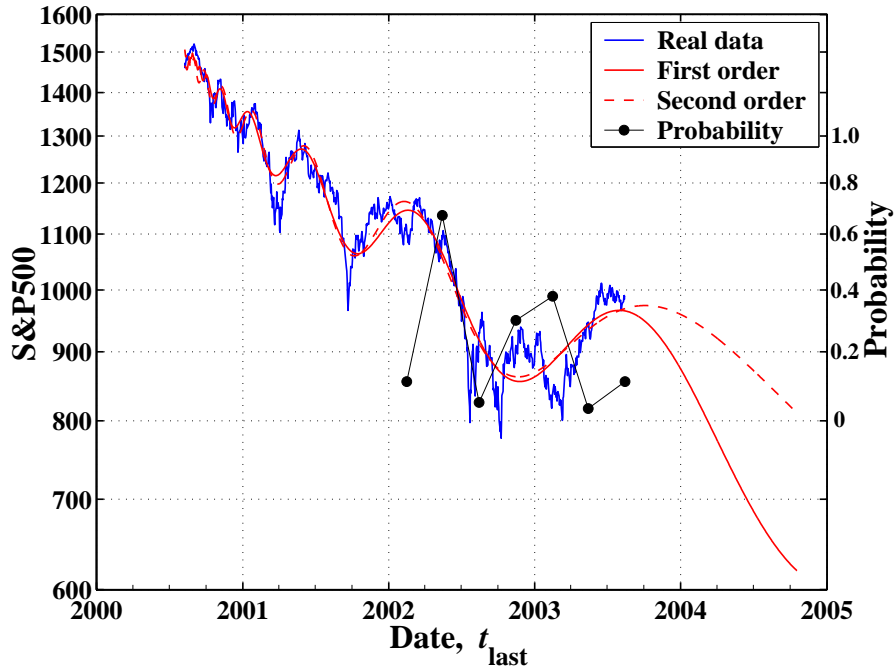}
\caption{\textbf{Left ordinate}: Fits of the S\&P
500 index over a time interval of three years
with a daily sampling rate using the first-order LPPL formulae
(\ref{Eq:lnpt}) and
the second-order LPPL formulae (\ref{Eq:Landau2}). The parameters
are the following: $t_c={\rm{2000/08/27}}$, $m=0.72$,
$\omega=9.2$, $\phi=4.62$, $A= 7.3123$, $B=-0.0037$, $C=-0.0008$,
and the r.m.s. of fit residuals is $\chi_1=0.03859$ for the first
order formula; and $t_c={\rm{2000/08/06}}$, $m=0.76$,
$\omega=11.4$, $\phi=1.03$, $\Delta_t= 2778$,
$\Delta_\omega=-22.6$, $A= 7.3245$, $B=-0.0031$, $C=-0.0007$, and
the r.m.s. of fit residuals is $\chi_2=0.03729$ for the second
order formula.  \textbf{Right
ordinate}: The probability that the simulated log-likelihood-ratio
exceeds the realized ratio as a function of $t_{\rm{last}}$.}
\label{Fig:LandauShift}
\end{figure}

We first present the fits of the S\&P 500 index time series from 2000/08/09 to
2003/08/15 with the first-order (\ref{Eq:lnpt}) and with the
second-order formula (\ref{Eq:Landau2}) in
Fig.~\ref{Fig:LandauShift}. We will extend below the upper time limit
of the fitting interval to provide further tests. The corresponding values
$\chi_1=0.03859$ and $\chi_2=0.03729$ give a log-likelihood ratio
$T = 51$. The probability that $T$ is greater than 51 for a
chi-square distribution with two degrees of freedom is $\approx
8\times 10^{-12}$, giving an extremely high confidence level
undistinguishable from $1$. It seems that the second-order formula
(\ref{Eq:Landau2}) is absolutely necessary. However, the rather
large value of the cross-over time $\Delta_t= 2778$ days = 7.6
years compared with the three year span of the time series
suggests that the transition from the first-order (\ref{Eq:lnpt})
and with the second-order formula (\ref{Eq:Landau2}) has just
begun in 2003/08/15. Redoing the calculation of the log-likelihood
ratio $T$ for
a shorter period also gives an extremely large confidence level,
which is suspect. For such a finite-time series, the validity of
the asymptotic Wilks test is questionable, especially in view of
the non-Gaussian and the large dependence in the residues of the
fits, which can be seen with the naked eye in
Fig.~\ref{Fig:LandauShift}. Wilks test assumes i.i.d. random
residues, which is certainly not the case at the daily scale. For
a weekly time step, the residues are less correlated. Redoing the
fits using a weekly time scale, we obtain $T=8.7$, giving a
confidence level of $98.7\%$. This enormous change in the
confidence level casts doubts on the validity of the Wilks test
and does not allow us to conclude from it that the second-order
formula is necessary. More generally, because the price time
series have a very complicated nature, applying classical
statistical tests (like the Wilks test) to such time series is
very dangerous. It is thus desirable to develop simple and robust
(multivariate) statistics (defined in a moving time window). This
paper is a first step in this direction.

To assess the statistical significance of the second-order
formula, we propose the following alternative algorithm which is tailored
to address the impact of the noise structure up to monthly time scales.
\begin{enumerate}
\item Starting the fit with the first-order formula, we
decompose the residues in segments of one-month duration.

\item We reshuffle the one-month intervals of the residuals at random.
Since there are about three years of data = 36 months, there are
$36!$ (factorial of $36$) ways of reshuffling the monthly residues.

\item We add the
reshuffled residues to the first-order formula, which provides
us with a noisy synthetic log-periodic time series.

\item We fit this synthetic time series with the first-
and with the second-order formulas and calculate the corresponding
log-likelihood ratio $T$ for these two fits for this realization.

\item We redo steps 2-4 one thousand times and count how many times the
value empirical value $T=51$ is exceeded.
\end{enumerate}
This algorithm is nothing but a bootstrap with noise realizations
generated from the real data. In this way, we keep the genuine
structure of the dependence of real prices up to the monthly
scale. This allows us to test how the empirical dependence
structure of prices up to one month scale may interfere with the
detection of log-periodicity and of its frequency shift. The
monthly scale is a compromise between having many statistical
realizations (favoring smaller time intervals) and keeping as much
as possible all the idiosyncratic textures of the price times
series that decorate the large scale log-periodicity.

We perform 1000 simulations for the S\&P 500 index time series
from 2000/08/09 to $t_{\rm{last}}=\rm{2003/08/15}$. The averages
and standard deviations of the parameters obtained from the fits
with Eq.~(\ref{Eq:lnpt}) are the following:
$\langle{t_c}\rangle=\rm{2000/08/19}\pm 19~\rm{days}$,
$\langle{m}\rangle=0.72\pm 0.10$, $\langle{\omega}\rangle=9.4\pm
0.5$, $\langle{\phi}\rangle=3.33\pm 1.69$,
$\langle{A}\rangle=7.32\pm 0.03$, $\langle{B}\rangle=-0.048\pm
0.32$, $\langle{C}\rangle=0.0010\pm 0.0006$,
$\langle{\chi_1}\rangle=0.0363\pm 0.0013$. The averages and
standard deviations of the parameters obtained from the fits with
Eq.~(\ref{Eq:Landau2}) are the following:
$\langle{t_c}\rangle=\rm{2000/08/14}\pm 19~{\rm{days}}$,
$\langle{m}\rangle=0.76\pm 0.11$, $\langle{\omega}\rangle=9.7\pm
1.7$, $\langle{\phi}\rangle=3.24\pm 2.08$,
$\langle{\Delta_t}\rangle=5889\pm 2674$ days,
$\langle{\Delta_{\omega}}\rangle=-3.7\pm 40$,
$\langle{A}\rangle=7.32\pm 0.03$, $\langle{B}\rangle=-0.0039\pm
0.0029$, $\langle{C}\rangle=0.0009\pm 0.0006$,
$\langle{\chi_2}\rangle=0.0358\pm 0.0014$. The average and
standard deviation of the log-likelihood ratio are $\langle T
\rangle = 22.7\pm 25.8$. The probability that $T>51$ is found to
be $11.1\%$. Thus, according to this bootstrap method, the null
hypothesis that S\&P 500 has not experienced a log-periodic
frequency shift cannot be rejected at a significance level of
10\%. However, the null hypothesis can be rejected at a
significance level of 12\%. Thus, the situation is less clear than
with the Wilks test which assumes asymptotic Gaussian i.i.d. noise
statistics but the evidence suggests that the transition to a
log-periodic frequency shift has started. This has an important
implication for the future evolution of the S\&P 500 index, as the
first-order (continuous line) and second-order (dashed line)
formulas diverge significantly after 2003/08/15, as shown in
Fig.~\ref{Fig:LandauShift}.

We have performed exactly the same procedure for other $t_{\rm
last}$ chosen earlier than 2003/08/15, from $t_{\rm last}=$2002/02/15 to
$t_{\rm{last}}=\rm{2003/08/15}$ in step of 3 months, giving a
total of 7 time periods. For each $t_{\rm{last}}$, we calculate
the empirical log-likelihood ratio and the associated probability
${\rm{Pr_{t_{\rm last}}}}$ that this ratio is exceeded, by using
the above bootstrap method. This gives:
${\rm{Pr_{2002/02/15}}}=11\%$, ${\rm{Pr_{2002/05/15}}}=67\%$,
${\rm{Pr_{2002/08/15}}}=5.1\%$, ${\rm{Pr_{2002/11/15}}}=30\%$,
${\rm{Pr_{2003/02/15}}}=38\%$, ${\rm{Pr_{2003/05/15}}}=3.4\%$, and
${\rm{Pr_{2003/08/15}}}=11\%$. The plot of
${\rm{Pr}}_{t_{\rm{last}}}$ as a function of ${t_{\rm{last}}}$ is
shown in Fig.~\ref{Fig:LandauShift}, with the scale indicated on
the right vertical ordinate. Overall, ${\rm{Pr_{t_{\rm last}}}}$
tends to decrease, which implies a progressive increasing
relevance of the second-order formula compared with the
first-order formula. The small value of ${\rm{Pr_{2002/02/15}}}$
is caused by the distortion of the prices with a local trough
around 2002/02/15, while that of ${\rm{Pr_{2002/08/15}}}$ results
from the local sharp peak around 2002/08/15 and probably the
preceding crash as well. This effect has been observed in
\cite{ZS03RG} (see Fig. 6 therein).

These results open seriously the possibility that the
S\&P 500 index has started to cross-over from the first-order to
the second-order formula already in 2003/08/15. The corresponding fits shown in
Fig.~\ref{Fig:LandauShift} suggests that there could be a delay in
the drop predicted in 2003/08/15 on the sole basis of the first-order formula
and perhaps a change of regime. In hindsight, we know now that the
change of regime turned out to be of a different more subtle nature, as we
discuss below.

As a word of caution, it is necessary to stress that the bootstrap
method is an in-sample method. In-sample results can differ
significantly from out-of-sample results because the bootstrap
method is performed under a fixed sample. In particular, it gives
conditional probabilities that converge to unconditional
probabilities only as the sample size tends to infinity. It is
difficult to assess a priori how close are bootstrap probabilities
to unconditional probabilities in our finite sample. This remains
a limitation of the approach.

\section{How to detect the end of the antibubble?}
\label{s1:EOA}

\subsection{Evolution of the fit parameters}
\label{s2:EvoFitPara}

According to standard economic theory, the prices of stocks must
reflect the discounted future capital flows. In practice, the
prices include the impact of news, from which anticipation on
future cash flows is made, as well as behavioral biases and
herding among investors. This provides a mixture of endogeneity
and exogeneity \cite{SorMalMuz,JS02XXX}. The antibubble phase is
supposed to reflect mostly the impact of the behavioral part which
leads to self-reinforced pessimism intermittently interrupted by
transient phases of optimism.

If the antibubble pattern is to be a correct description of the
market prices, a necessary condition is that its parameters should
be robust, that is, approximately constant as a function of time.
On the other hand, as time flows, the cumulative effect of
exogenous news may detune progressively the antibubble pattern.
This phenomenon may be accelerated in the presence of a strong
exogenous shock. One can thus view the unfolding of an antibubble
as a dynamic process with competing forces attempting to maintain
and to destroy the LPPL structure.

\begin{figure}
\centering
\includegraphics[width=12cm]{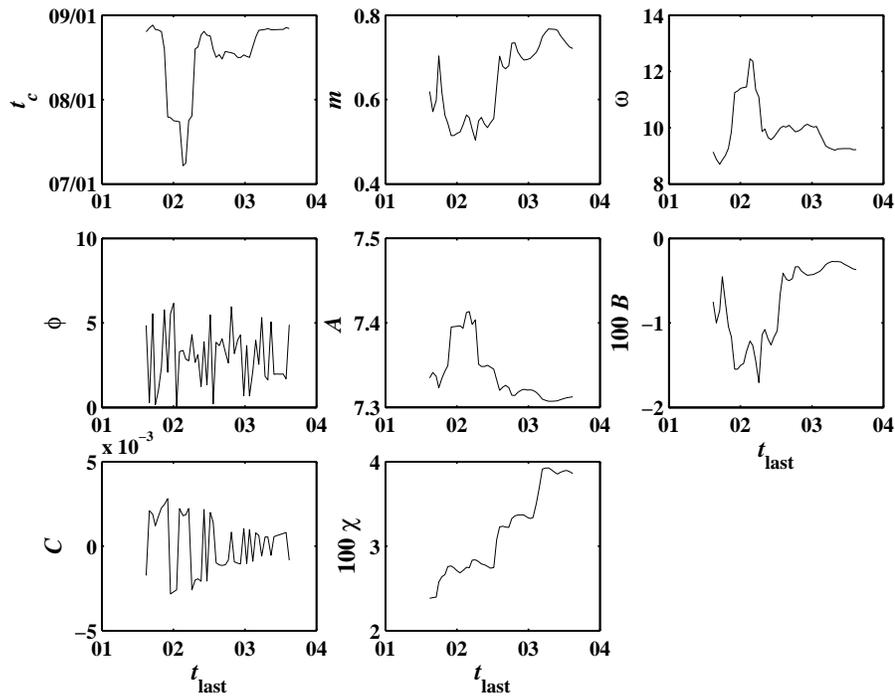}
\caption{Evolution of the parameters of the fit of the
S\&P 500 with the log-periodic formula (\ref{Eq:lnpt})
as a function of time. Note that the tick labels on
the ordinate of the left-top plot is ``mm/dd'' for year 2000
whereas the labels on the ticks of the abscissa refer to the last two digits
of the years. Each tick is positioned of 1st January.}
\label{Fig:EOA_PE}
\end{figure}

We fit the US S\&P 500 index to the LPPL formulae (\ref{Eq:lnpt})
over a running window from 2000/08/09 to
$t_{\rm{last}}$, where $t_{\rm{last}}$ is sampled at a bi-weekly rate in
the interval from 2001/08/15 to 2003/08/15. Figure \ref{Fig:EOA_PE}
shows the evolution of the fit parameters $t_c$, $m$, $\omega$,
$\phi$, $A$, $B$, $C$ and of the root-mean-square (r.m.s.) of the fit
residuals $\chi$.
The most noticeable structure in these plots is the deviation of the
parameters from their approximately constant value, which occurred at the
end of 2001 and lasted one to two quarters. This deviation is associated
with the ``crash'' of August 2001 \cite{ZS03RG}. Notice that the r.m.s. $\chi$
has been growing in steps, each step corresponding roughly to the
pronounced drops
and associated volatility at successive bottoms of the log-periodic trajectory.

Based on Figure \ref{Fig:EOA_PE}, there does not seem to be a flagrant
change of regime up to the most recent investigated
$t_{\rm{last}}=$2003/08/15, so that other tests are needed.

\subsection{Construction of scenarios with uncontaminated reference}
\label{s2:ED}

To quantify the possibility that the antibubble may have disappeared
or will disappear,
we construct two classes of scenarios and test how the LPPL fits distinguish
between them. Consider the S\&P 500 from the onset of the
antibubble (approximately 2000/08/09) to a time
$t_{\rm{last}}$. The scenarios are obtained by extending this time series
for six months after $t_{\rm{last}}$ by
\newcounter{Lcount}
\begin{list}{Class \Roman{Lcount}:}
     {\usecounter{Lcount} \setlength{\rightmargin}{\leftmargin}}
     \item continuing the log-periodic formula with noise added to it,
\label{ts1}
     \item performing a random walk with daily
volatility equal to the historical volatility over the same period. \label{ts2}
\end{list}
Class I corresponds to the continuation of the antibubble regime.
Class II corresponds to a regime switch at $t_{\rm{last}}$ from
the antibubble to a structureless price trajectory.

We generate $N=1000$ time series for each class and then fit each
of them by the LPPL formula (\ref{Eq:lnpt}). In the simulations,
we use noise generated by a GARCH (generalized auto-regressive
conditional heteroskedasticity) model, which is a process often
taken as a benchmark in the financial industry and which takes
into account volatility persistence. The innovations of the GARCH
noise process have been drawn from a Student distribution with $3$
degrees of freedom with a variance equal to that of the fit
residuals of the real data. This ensures a reasonable
correspondence between the statistical properties of these
synthetic time series and the known properties of the empirical
distribution of returns.

Calling ${\vec X}$ the vector of parameters $t_c$, $m$,
$\omega$, $\phi$, $A$, $B$, $C$, and $\chi$,
we thus obtain two sets of $N$ vectors for each class. The gist of this
test is to quantify the differences in the distributions of the parameters
${\vec X}$ in the two classes: if the differences are significant,
this procedure provides a natural classification to apply to the
real realization in order to decide whether it belongs to Class I or Class II.
Specifically, if the antibubble indeed continues up to $t_{\rm{last}}+6$ months
with a price
trajectory close to the extrapolation of the log-periodic fit performed
up to $t_{\rm{last}}$, one could expect that the parameters of the fits
of the time series up to $t_{\rm{last}}+6$ months
with the LPPL formula (\ref{Eq:lnpt}) should be close to the set found
for Class I and far from those found for Class II. Conversely, if the
S\&P 500 index switches to a random walk after $t_{\rm{last}}$, one should find
the corresponding parameters of the log-periodic fit
to depart from the set found
for Class I while being compatible with those found for Class II. This test
is part of a large class of pattern recognition methods
\cite{Gelfand76,KBbook}.
Using the pattern recognition language, we refer to
each time series as an object to be classified (either in Class I or Class II).

\begin{figure}
\centering
\includegraphics[width=12cm]{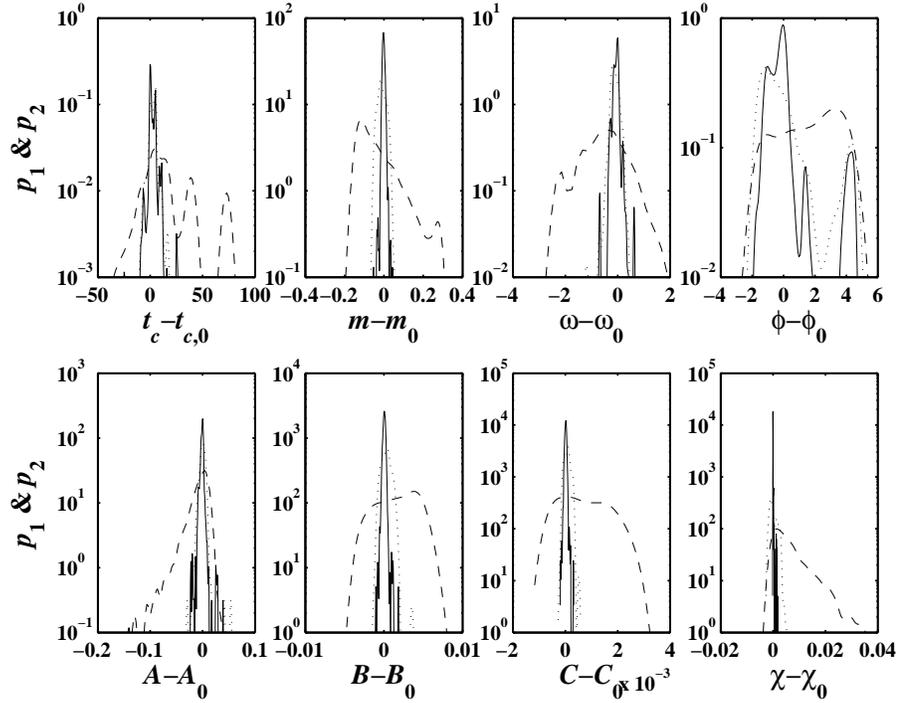}
\caption{Probability density functions $p_1(x-x_0)$ (continuous
lines), $p_2(x-x_0)$ (dashed lines), and $p^*_1(x-x_0)$ (dotted
lines) associated with a reference antibubble from 2000/08/09 to
$t_{\rm{last}} ={\rm{2003/08/15}}$ for the eight parameters deriving from
the fits with the log-periodic formula (\ref{Eq:lnpt}). $p^*_1(x-x_0)$
is modified from $p_1(x-x_0)$ by using a more realistic noise process,
as explained in the text.}
\label{Fig:EOA:pdf}
\end{figure}

Figure \ref{Fig:EOA:pdf} plots the probability density functions
(PDFs), $p_1(x-x_0)$ (solid lines) and $p_2(x-x_0)$ (dashed
lines), of the difference between a given fit parameter $x$ and
its reference value $x_0$, for the two classes associated with the
antibubble that developed from 2000/08/09 to $t_{\rm{last}}
={\rm{2003/08/15}}$. The index $1$ (respectively $2$) refers to
Class I (respectively II). The variable $X$ stands for any of the
parameters $t_c$, $m$, $\omega$, $\phi$, $A$, $B$, $C$, and
$\chi$. The reference value $x_0$ is the value of the parameter
obtained in the fit of the antibubble from 2000/08/09 to
$t_{\rm{last}} ={\rm{2003/08/15}}$ with the log-periodic formula.
The differences between each pair of PDFs are significant:
$p_1(x-x_0)$ concentrates around $x-x_0=0$, as could be expected,
while $p_2(x-x_0)$ exhibits a much larger dispersion with much
slower decaying tails.

In pattern recognition methods, it is necessary to
define two types of errors that can occur in a classification scheme
using a given fit parameter $x$.  An error of type I
occurs when the hypothesis, which is true, is rejected
(a ``false negative'' in terms of null hypothesis testing). Errors of type I
occur with a complementary cumulative
probability $P_1(x)$ measured as the proportion of the objects in class I
with a deviation $|X-x_0|$ greater than $|x-x_0|$:
\begin{equation}
P_1(x) = \lim_{N\to\infty}\frac{\sharp\{X:|X-x_0|>|x-x_0|~\&~X \in
{\rm{I}}\}}{N}~, \label{Eq:P1x}
\end{equation}
where $\sharp$ is the operator counting the number of elements in a
given set. An error of type II occurs
when an hypothesis, which is false, is accepted (a ``false positive'' or
``false alarm'' in
terms of null hypothesis testing). Errors of type II
occur with a cumulative probability  $P_2(x)$ measured as the
proportion of the objects in class II with the deviation $|X-x_0|$
smaller than $|x-x_0|$:
\begin{equation}
P_2(x) = \lim_{N\to\infty}\frac{\sharp\{X:|X-x_0|<|x-x_0|\}~\&~X
\in {\rm{II}}\}}{N}~. \label{Eq:P2x}
\end{equation}
By definition, $\lim_{x\to x_0}P_1(x)=1$, $\lim_{x\to
x_0}P_2(x)=0$, $\lim_{|x-x_0|\to\infty}P_1(x)=0$, and
$\lim_{|x-x_0|\to \infty}P_2(x)=1$.

\begin{figure}
\centering
\includegraphics[width=12cm]{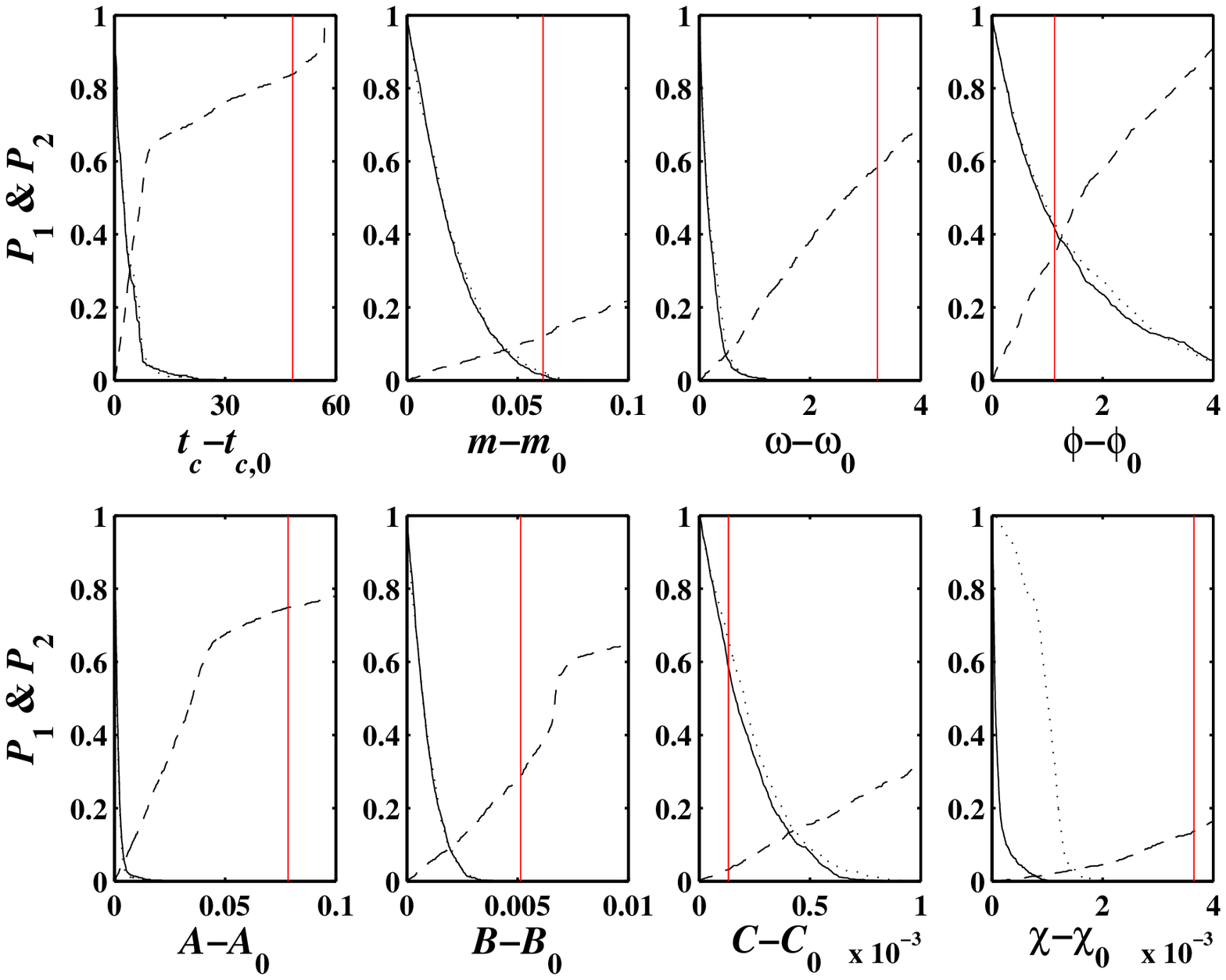}
\caption{Probabilities $P_1(x)$ (continuous lines), $P^*_1(x)$
(dotted lines), and $P_2(x)$ (dashed lines) corresponding to the
reference antibubble from 2000/08/09 to $t_{\rm{last}}
={\rm{2001/08/15}}$ as functions of eight parameters derived from the
fits with the log-periodic formula (\ref{Eq:lnpt}). The
vertical lines indicate the realized values of $|x-x_0|$. }
\label{Fig:EOA:P1P2:1}
\end{figure}

\begin{figure}
\centering
\includegraphics[width=12cm]{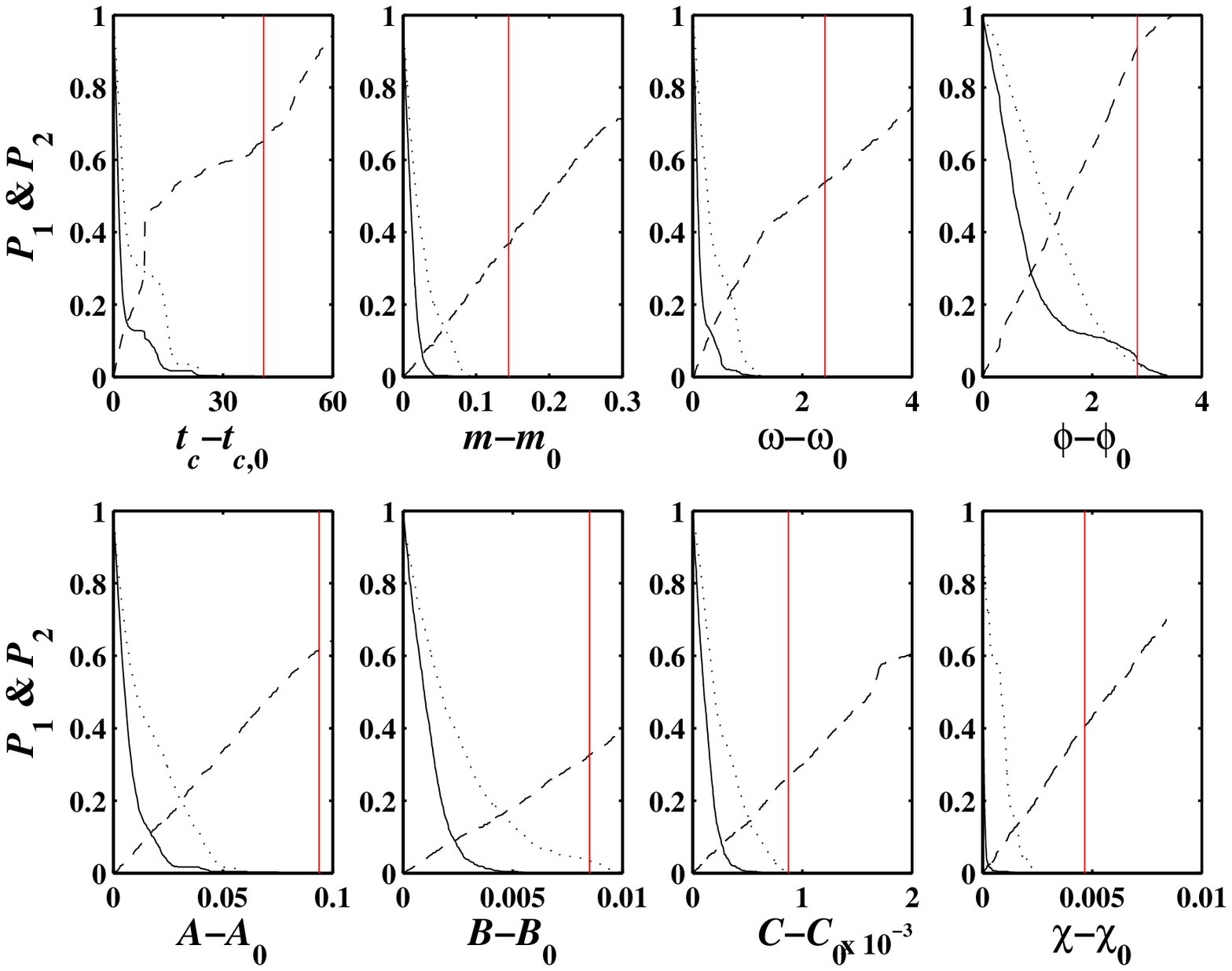}
\caption{Same as Fig.~\ref{Fig:EOA:P1P2:1} for $t_{\rm{last}}
={\rm{2002/02/15}}$.} \label{Fig:EOA:P1P2:2}
\end{figure}

\begin{figure}
\centering
\includegraphics[width=12cm]{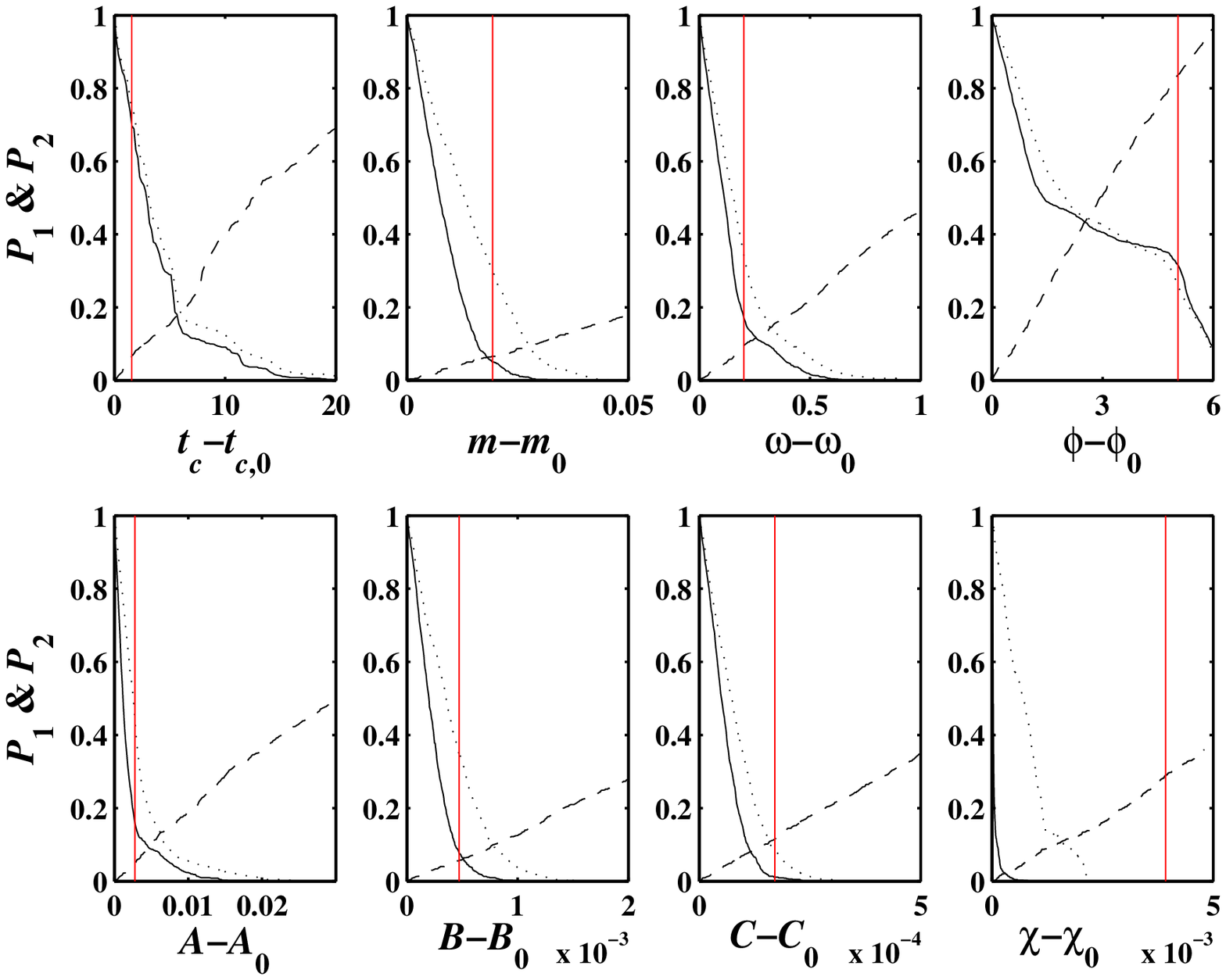}
\caption{Same as Fig.~\ref{Fig:EOA:P1P2:1} for $t_{\rm{last}}
={\rm{2002/08/15}}$.} \label{Fig:EOA:P1P2:3}
\end{figure}

\begin{figure}
\centering
\includegraphics[width=12cm]{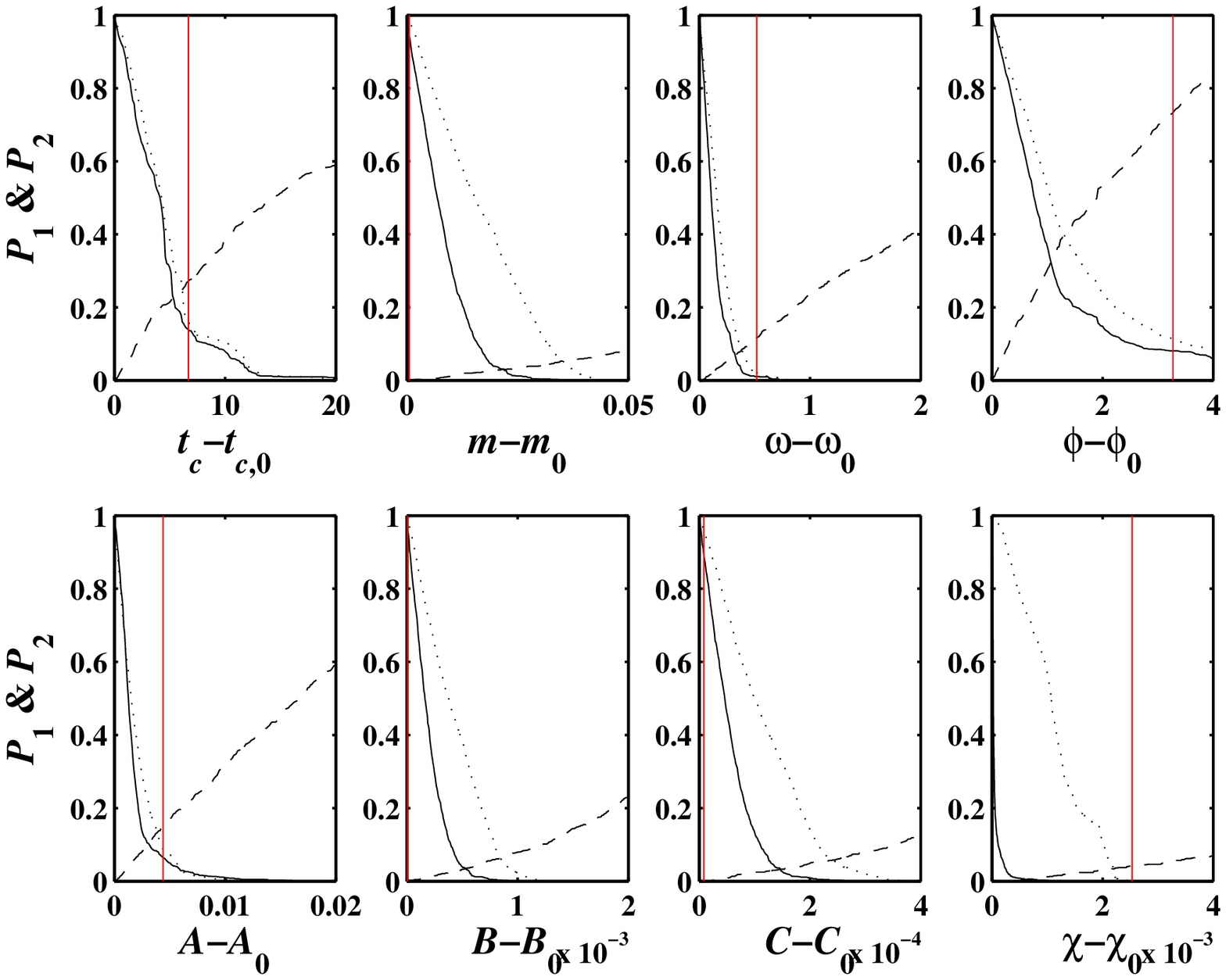}
\caption{Same as Fig.~\ref{Fig:EOA:P1P2:1} for $t_{\rm{last}}
={\rm{2003/02/15}}$.} \label{Fig:EOA:P1P2:4}
\end{figure}

\begin{figure}
\centering
\includegraphics[width=12cm]{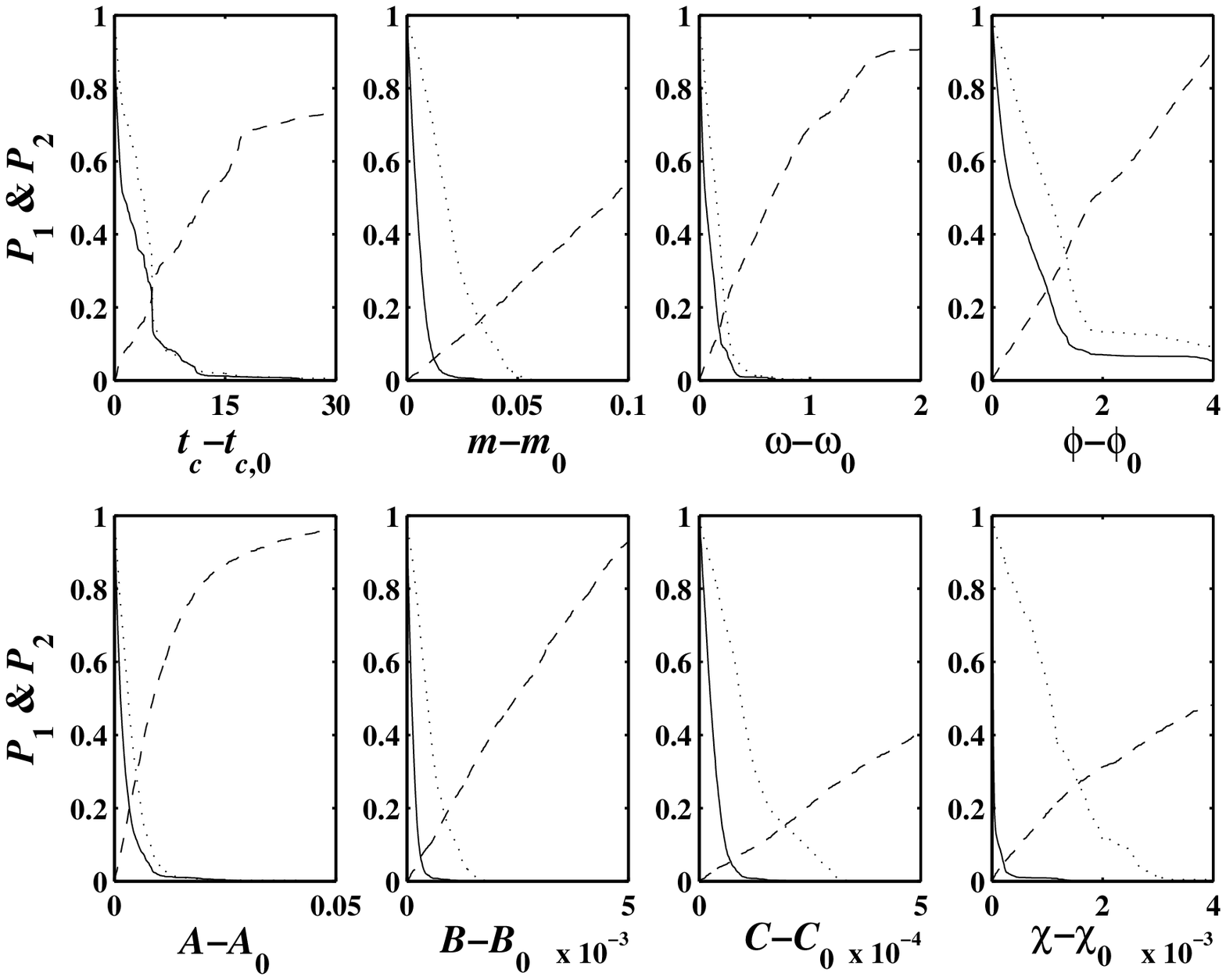}
\caption{Same as Fig.~\ref{Fig:EOA:P1P2:1} for $t_{\rm{last}}
={\rm{2003/08/15}}$.} \label{Fig:EOA:P1P2:5}
\end{figure}

Figure \ref{Fig:EOA:P1P2:1} shows the probabilities $P_1(x)$ and
$P_2(x)$ constructed by taking as the reference the antibubble on
the S\&P 500 from 2000/08/09 to
$t_{\rm{last}} ={\rm{2001/08/15}}$, for the seven fit
parameters and for the r.m.s $\chi$.
Figures \ref{Fig:EOA:P1P2:2}
to \ref{Fig:EOA:P1P2:5} are the same for $t_{\rm{last}}
={\rm{2002/02/15}}$, ${\rm{2002/08/15}}$, ${\rm{2003/02/15}}$, and
${\rm{2003/08/15}}$, respectively.

As seen in Figure \ref{Fig:EOA:pdf}, the PDF's for Class I are
extremely narrow. One may wonder if this is not due to our use of
the GARCH  process which gives a too conservative estimate of the
noise impact. To test this possibility, we use another noise
generating process. Rather than generating noise synthetically, we
construct the time series of the residues $\epsilon(t)$ obtained
from the log-periodic fit of the reference time series with the
LPPL formulae. We then extract at random a six month segment of
$\epsilon(t)$ which is the noise taken to decorate the extended
series from $t_{\rm{last}}$ to $t_{\rm{last}} +6$ months for Class
I objects. Having thus generated new objects of Class I, we
calculate the new PDFs and the probabilities $P^*_1(x)$ defined as
the proportion of the objects (with ``residual'' noise) in class I
with a deviation $|X-x_0|$ greater than $|x-x_0|$. The new PDFs
$p^*_1(x-x_0)$ are shown as dotted line in Figure
\ref{Fig:EOA:pdf}. The dependence of the corresponding
$P^*_1(x)$ (defined as $P_1(x)$ but using $p^*_1(x-x_0)$
instead of $p_1(x-x_0)$) for the seven
parameters and for the r.m.s. as a function of $|x-x_0|$ are shown
as the dotted lines in Figs.~\ref{Fig:EOA:P1P2:1} to
\ref{Fig:EOA:P1P2:5}. As expected, using past realized residuals
gives slightly larger dispersions but the differences are not
large. This confirms the large difference between objects in Class
I and in Class II.

To qualify the continuation of the antibubble based
on the measured value $x$ of one parameter, one would like to have
both $P_1(x)$ large (above some threshold) and $P_2(x)$ small
(below some threshold). The first condition ($P_1(x)$ sufficiently
large) tells us that the realized deviation is well within the
normal fluctuations of objects in Class I. The second condition
($P_2(x)$ sufficiently small) indicates that it is improbable to
obtain such a value of $x$ if the price series was not an
antibubble. These two conditions quantify how much deviation of
$x$ from the reference value $x_0$ is tolerable to qualify the
additional six month of data as a continuation of the antibubble.

In practice, using $t_{\rm{last}} ={\rm{2003/08/15}}$, one has to wait
an additional 6 month and analyze the realized time series as an object
to be classified according to the above scheme. To test the sensitivity
and reliability of this procedure, it is natural to turn to data in the
past of $t_{\rm{last}} ={\rm{2003/08/15}}$
to simulate how this method would have worked in this past. We will then
turn below to examine the data posterior to $t_{\rm{last}} ={\rm{2003/08/15}}$.

\subsection{Ex-post tests}
\label{s2:CP}

To assess the validity of the proposed method, we test it
retroactively. The test consists in taking the price time series
from 2000/08/09 to $t_{\rm{last}}$ as the reference and in
applying the procedure described in section \ref{s2:ED} for each
$t_{\rm{last}}$, with $t_{\rm{last}}$ taking the values
${\rm{2001/08/15}}$, ${\rm{2002/02/15}}$, ${\rm{2002/08/15}}$, and
${\rm{2003/02/15}}$, with a time step of six months.
We use the realized values of the fitted
parameters obtained for the time series extending to
$t_{\rm{last}}+6$ months to obtain the two probabilities $P_1$ and
$P_2$. The realized values of $P_1$, $P_1^*$ and $P_2$ are listed
in Table \ref{Tb1}. The realized values of the fit parameters for
the time series extending to $t_{\rm{last}}+6$ months are
indicated by the vertical line in Figs.~\ref{Fig:EOA:P1P2:1} to
\ref{Fig:EOA:P1P2:5}.

The rather poor results (small $P_1$'s, $P_1^*$'s and large
$P_2$'s) for the two earlier times
$t_{\rm{last}}={\rm{2001/08/15}}$ and
$t_{\rm{last}}={\rm{2002/02/15}}$ can probably be attributed to
the fact that the log-periodic structure was not yet sufficiently
developed and was dominated by noise. A large $P_2$ in particular means
that the six-month extension from $t_{\rm{last}}$ to
$t_{\rm{last}}+6$ months had similarity with a random walk.
For the two later times $t_{\rm{last}}={\rm{2002/08/15}}$ and
$t_{\rm{last}}={\rm{2003/02/15}}$, we observe often large $P_1$'s
and small $P_2$'s, suggesting that the antibubble has continued to
develop. We should also stress that all parameters are not
equivalent for the decision process. For instance, the r.m.s. of
fit residuals is almost insensitive to the phase $\phi$, which
explains why the values of $P_1$ and $P_2$ are completely
uninformative for the phase.

\subsection{Impact of past regime switching: contaminated reference}
\label{s2:PastDetune}

The previous tests have been performed with the hypothesis that
the antibubble have been genuinely continuing until
$t_{\rm{last}}$. This condition has allowed us to take the
parameters of the fits of the time series up to $t_{\rm{last}}$ as
references. But what about the possibility that the price time
series has already switched to a random walk? It could be the case
that we may incorrectly believe in the antibubble continuation
until $t_{\rm{last}}$ while in fact a part of the past time series
is already in the random walk regime. The reference values of the
fitted parameters would then be incorrect, leading to possible
distortions in the calculation of $P_1$ and $P_2$.

We thus also need to take into account the fact that the regime switch may
have happened in the past, to quantify what is its effect in qualifying
its future. To address this question, we replace the data of
the last six months of the reference series ending at
$t_{\rm{last}}$ by a random walk with time steps
equal to the historical volatility. Specifically, from the beginning
of the time series to $t_{\rm{last}}-6$ months, the time series is
the S\&P 500 data. From $t_{\rm{last}}-6$ months to $t_{\rm{last}}$,
we extend the S\&P 500 data by generating a random walk. The resulting
time series ending at $t_{\rm{last}}$ is then taken at the believe-to-be-true
antibubble to which we apply the above procedure described in section
\ref{s2:ED}.

\begin{figure}
\centering
\includegraphics[width=12cm]{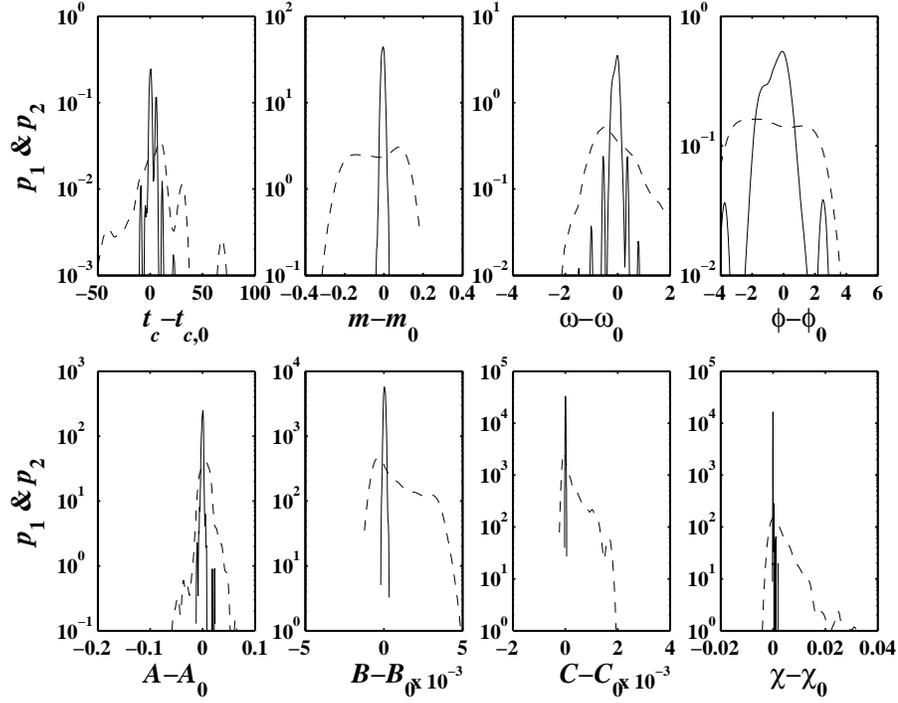}
\caption{Probability density functions $p_1(x-x_0)$ (continuous
lines) and $p_2(x-x_0)$ (dashed lines) associated with a
contaminated reference antibubble from 2000/08/09 to
$t_{\rm{last}} ={\rm{2003/08/15}}$ for the eight parameters
derived from the
fits with the log-periodic formula (\ref{Eq:lnpt}).
See Sec.~\ref{s2:PastDetune} for details.} \label{Fig:EOA2:pdf:5}
\end{figure}

\begin{figure}
\centering
\includegraphics[width=12cm]{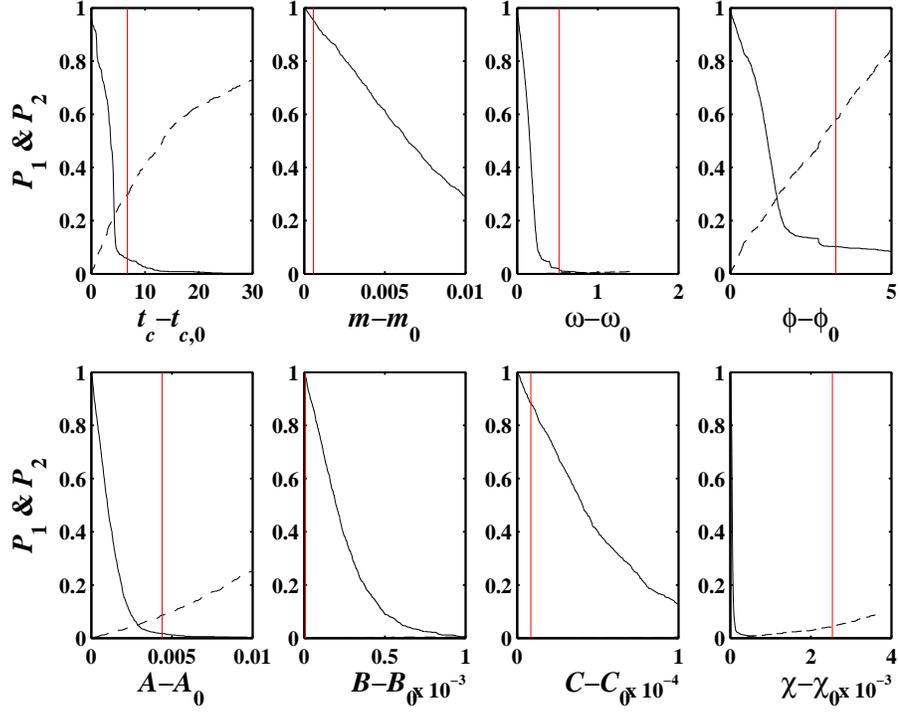}
\caption{Probabilities $P_1(x)$ (continuous lines) and $P_2(x)$
(dashed lines) corresponding to a contaminated reference antibubble
from 2000/08/09 to $t_{\rm{last}} ={\rm{2001/08/15}}$ as functions
of the eight parameters derived from the
fits with the log-periodic formula (\ref{Eq:lnpt}).
The vertical lines indicate the realized
values of $|x-x_0|$.} \label{Fig:EOA2_P1P2_4}
\end{figure}

\begin{figure}
\centering
\includegraphics[width=12cm]{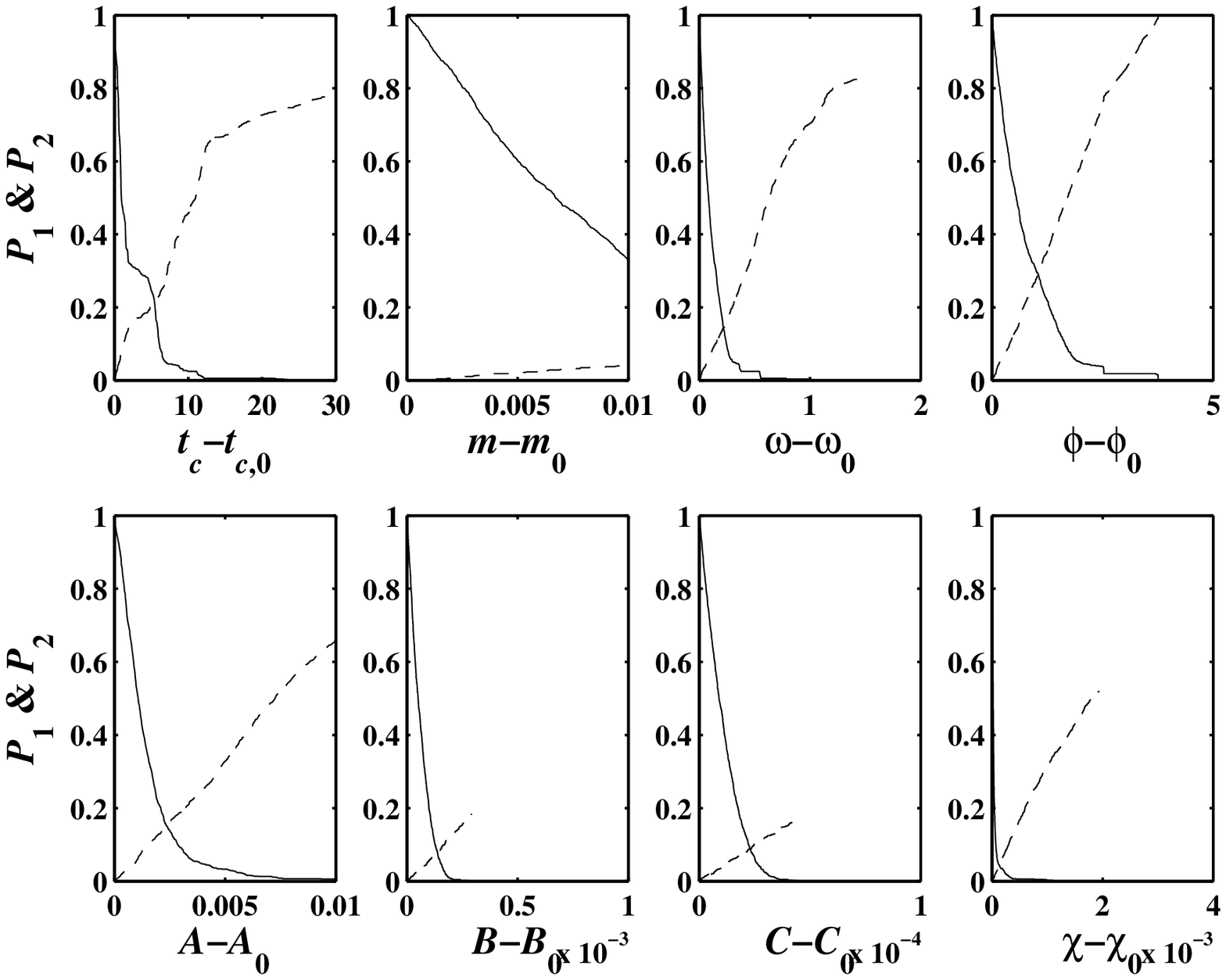}
\caption{Same as Fig.~\ref{Fig:EOA2_P1P2_4} for $t_{\rm{last}}
={\rm{2003/08/15}}$.} \label{Fig:EOA2_P1P2_5}
\end{figure}

The tests are performed for $t_{\rm{last}} = {\rm{2003/02/15}}$
and ${\rm{2003/08/15}}$. The results are given in
Figs.~\ref{Fig:EOA2:pdf:5} to \ref{Fig:EOA2_P1P2_5}, where the
vertical lines indicate the values of the realized $|x-x_0|$
for the time series ending at $t_{\rm{last}}+6$ months.
In Fig. \ref{Fig:EOA2:pdf:5}, one observes a broadening of $p_1$ as
can be expected.

\begin{figure}
\centering
\includegraphics[width=12cm]{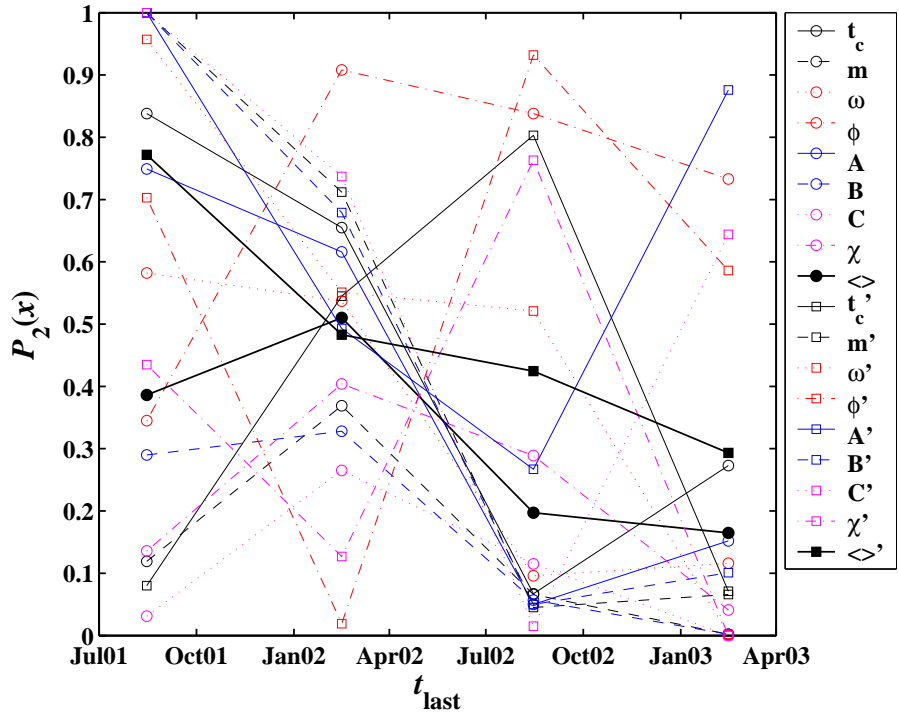}
\caption{Probability $P_2(x)$ as a function of $t_{\rm{last}}$ for the
different parameters. The non-prime
parameters are defined in Sec.~\ref{s2:ED} and correspond to the uncontaminated
reference case. The primed parameters are defined in Sec.~\ref{s2:PastDetune}
and correspond to the
contaminated reference case. The thick lines with close circles
and squares give the average over the eight parameters for the two
types of $P_2$ (uncontaminated and contaminated respectively).}
\label{Fig:EOA:P2}
\end{figure}

Fig.~\ref{Fig:EOA:P2} shows $P_2$ as a function of $t_{\rm{last}}$
for the uncontaminated cases (circles) studied in Sec.~\ref{s2:ED}
and for the contaminated cases (squares) studied in this section
for each of the 8 parameters. The adjective uncontaminated
(respectively contaminated) refers to taking the true time series
up to $t_{\rm{last}}$ (respectively to replacing the true time
series by a random walk in the interval from $t_{\rm{last}}-6$
months to $t_{\rm{last}}$). The overall picture is that the
squares for the contaminated case tend to spread more uniformly in
$[0,1]$ while the uncontaminated case becomes more concentrated
towards smaller values of $P_2$ for the last three $t_{\rm last}$.
The means of $P_2$ for the uncontaminated (respectively
contaminated) cases are shown in thick lines with closed circles
and squares respectively. One can observe that the $P_2$'s for the
contaminated case are significantly larger than for the
uncontaminated case, suggesting that it may be possible to
distinguish between them.

\begin{figure}
\centering
\includegraphics[width=12cm]{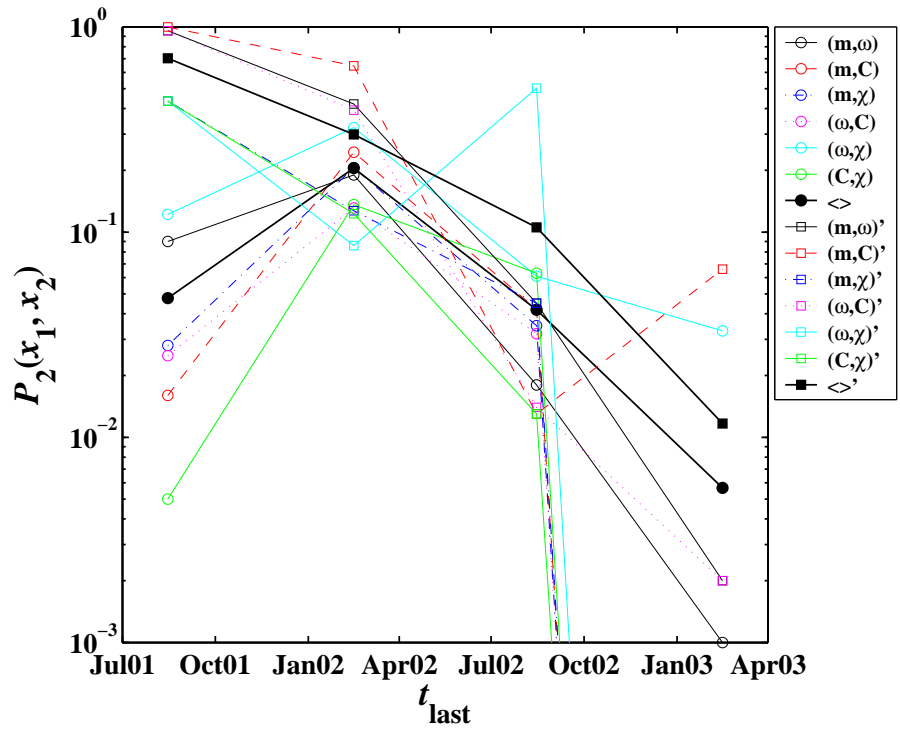}
\caption{Combined probability $P_2(x_1,x_2)$ as a function of
$t_{\rm{last}}$ for both uncontaminated (circles) and contaminated
(squares) cases.
Pairs $(x_1,x_2)$ are formed by taking $x_1$ and $x_2$ in the set
$\{m$, $\omega$, $C$, $\chi \}$. The thick lines with close
circles and squares give the average over the eight parameters for
the two types of $P_2$ (uncontaminated and contaminated respectively).}
\label{Fig:EOA:6P2}
\end{figure}

Fig.~\ref{Fig:EOA:6P2} plots the product of two $P_2$'s associated
with two parameters chosen from the set $\{m$, $\omega$, $C$,
$\chi \}$, both for uncontaminated (circles) and contaminated
(squares) cases. We denote $P_2(x_1,x_2)$ the product of the two
$P_2$'s associated with the parameters $x_1$ and $x_2$. We also
perform the averages over all pairs of $P_2(x_1,x_2)$, for the
uncontaminated and for the contaminated cases, which are shown
with filled circles and squares respectively. The construction of
$P_2(x_1,x_2)$ constitutes one step in the direction of a decision
for the qualification or disqualification of the antibubble which
should be ideally performed on the basis of the full multivariate
distributions over all parameters simultaneously. One can observe
significantly larger $P_2(x_1,x_2)$'s for the uncontaminated cases
compared with the contaminated cases. The fact that $P_2(x_1,x_2)$
tends to decrease for the last three points can be interpreted as
follows: the data has accumulated more so that the log-periodic
structure has become more developed, which constrains more the
fits. As a consequence, it is thus less probable to misinterpret a
random walk for a genuine LPPL antibubble between $t_{\rm{last}}$
and $t_{\rm{last}}+6$ months. Notice also the slower decay of
$P_2(x_1,x_2)$ for the last two points for the uncontaminated
cases compared with the contaminated cases.

\subsection{Testing the end of the antibubble: formulation and implementation}

As an empirical implementation of our detection method,
we propose the following test for the possible end of
the antibubble, based on selected scenarios
for the future. To illustrate the method, we take the date
of 2003/08/15 as the end of the known time series, and then project
several possible scenarios over the following
six months. For each scenario, the characteristic probabilities
$P_1$ and $P_2$ are calculated and used to characterize the
two possible outcomes: (i) the antibubble continues or (ii) the antibubble
has ended. We then apply this procedure to the realized data from
2003/08/15 to 2004/02/15 (2003/08/15 $+$ six months). In the first version
of this paper available in August 2003 (v.1 at
http://arxiv.org/abs/cond-mat/0310092),
we performed the first part  in real time and out-of-sample. The time
elapsed since allows us to describe the conclusion of this test on
the realized data.

\subsubsection{Synthetic scenarios \label{s2:ScenariosI}}

Let us thus consider 2003/08/15 (for which the S\&P 500 was slightly
below 1000)
as the date from which we project scenarios
to test for the continuation or ending of the antibubble. We extend the
price time series beyond 2003/08/15 by constructing
seven different scenarios of the future S\&P 500
evolution for the next six months:
\begin{itemize}
\item[(i)] a random walk taking the
S\&P 500 to the value 1200;

\item[(ii)] a random walk taking the S\&P 500
to 1100;

\item[(iii)] a random walk taking the S\&P 500 to 1000;

\item[(iv)] a random walk taking the S\&P 500 to 900;

\item[(v)] a random walk taking the S\&P 500 to 800;

\item[(vi)] a continuation of the antibubble with
noise obtained by a GARCH process described in Sec.~\ref{s2:ED}
(Class I and $P_1$); and

\item[(vii)] a continuation of the antibubble
with noise obtained by drawing at random the residuals over six
previous months as in Sec.~\ref{s2:ED} (Class I and $P^*_1$).
\end{itemize}

We have generated 424 realizations for each of these seven
scenarios ending at 2004/02/15 (2003/08/15 plus 6 months). Each
realization, which has been fitted by the LPPL formula
(\ref{Eq:lnpt}), yields 7 parameters and the r.m.s.
For each realization, the
two probabilities $P_1$ and $P_2$ defined in (\ref{Eq:P1x}) and (\ref{Eq:P2x})
are obtained for the seven
parameters, from which their average and standard deviations are determined.
The results are shown in Table \ref{Tb2}. The most
striking observation is that $P_1$ is small (respectively large)
for the five random walk scenarios (respectively for the
continuation of the antibubble), while $P_2$ is large
(respectively small) for the five random walk scenarios
(respectively for the continuation of the antibubble). As
expected, for the five random walk scenarios, $P_1$ increases and
$P_2$ decreases with decreasing ending value of the synthetic
values. These results suggest that one should be able to
distinguish clearly the continuation of the antibubble from a
regime switch to a random walk beyond 2003/08/15. However, one should
keep in mind that
the real future evolution might be more complicated than a random
walk trajectory with consequences for the test which are
difficult to foresee.

The fact that $P_1$ is so small for the random walk scenarios
(i)-(v) and quite large for the continuation of the antibubble
scenarios (vi) and (vii) tells us something important. Recall that
$P_1$ quantifies the probability that the deviations on the LPPL
parameters is larger in a true LPPL antibubble continuation than
those obtained from the scenarios. Small $P_1$'s for the random
walk scenarios (i)-(v) means that, conditioned on the fact that we
believe (erroneously) that the scenarios (i)-(v) are genuine LPPL
structures, essentially any random realization decorating a true
LPPL structure would continue to qualify as a genuine LPPL
structure. In other words, $P_1$ can be interpreted as the
probability of existence of the LPPL antibubble. It is very small
for the random walk scenarios (i)-(v) and quite large for the
continuation of the antibubble scenarios (vi) and (vii).
Reciprocally, the fact that $P_2$ is so large for the random walk
scenarios  (i)-(v) is in line with the fact that LPPL fits give
large errors for these scenarios and thus, conditioned on the fact
that these scenarios are believed a priori to be genuine LPPL
antibubbles, it is very probable that random walk realizations
would give similar or even better LPPL fits. In other words, what
this test tells us is that, starting with a bad fit, additional
noise can give similar or better fits. In contrast, the low value
of $P_2$ for the continuation of the antibubble scenario (vi) and
(vii) means that random walk extensions are very unlikely to give
qualities of fits similar to those obtained on average for these
scenarios (vi) and (vii). In sum, the small $P_1$ and large $P_2$
found for the random walk scenarios (i)-(v) are good signals of
the end of the LPPL antibubble. In contrast,
the large $P_1$ and small $P_2$ found for the continuation scenarios (vi)
and (vii) are good signals of the continuation of the LPPL
antibubble.

Thus, we conclude that, given the present price
pattern, there is only a small probability of making an error in
diagnosis: (a) if we obtain a small $P_1$ and a large $P_2$ in the
realized six months from 2003/08/15 to 2004/02/15, we will
conclude that the antibububble has ended; (b) in contrast, if
we obtain a large $P_1$ and a small $P_2$, we will conclude that
the antibubble continues.

\subsubsection{Realized probabilities and apparent end of the US antibubble}
\label{s2:ScenariosII}

Let us apply the test just described to the realized data from
2003/08/15 to 2004/02/15. In the first version of our paper presented in
Sec.~\ref{s2:ScenariosI}, we could not conclude that the
antibubble had ended yet and suggested that we would be able to decide
when the data till Feb 2004 would become available. Here we complete this test.

\begin{figure}
\centering
\includegraphics[width=12cm]{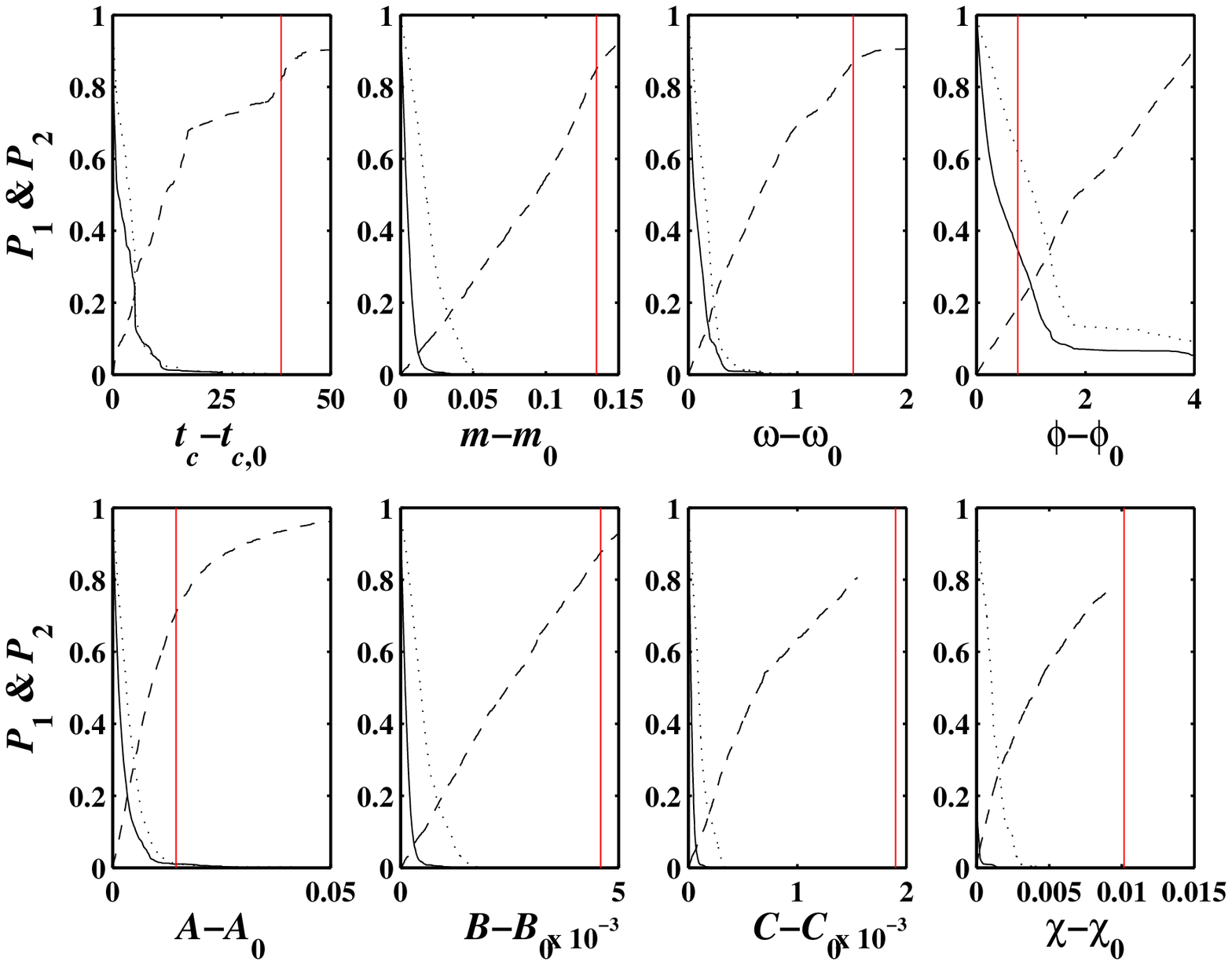}
\caption{Probabilities $P_1(x)$ (continuous lines), $P^*_1(x)$
(dotted lines), and $P_2(x)$ (dashed lines) corresponding to the
reference antibubble from 2000/08/09 to $t_{\rm{last}}
={\rm{2004/02/15}}$ as functions of eight parameters derived from
the fits with the log-periodic formula (\ref{Eq:lnpt}). The
vertical lines indicate the realized values of $|x-x_0|$.}
\label{Fig:EOA:P1P2:5bis}
\end{figure}

For this, Figure \ref{Fig:EOA:P1P2:5bis} shows the probabilities $P_1$
(continuous lines), $P_1^*$ (dotted lines), and $P_2$ (dashed
lines) corresponding to the reference antibubble from 2000/08/09
to 2003/08/15 as functions of eight parameters derived from the
fits with the first-order log-periodic formula, which was shown in
Fig.~\ref{Fig:EOA2_P1P2_5}. The vertical lines indicate the
realized values of $|x-x_0|$, where $x_0$ is the reference value.
One can see that the $P_1$'s are very small and the $P_2$'s are very large
for all parameters but the phase $\phi$, which was previously shown to be
irrelevant anyway. The small values of $P_1$ and large values of $P_2$
indicate that the antibubble in the USA has apparently ended.

\subsection{S\&P 500 in other currencies}
\label{s2:ForeignCurrencies}

In the previous tests, the S\&P 500 index was valued in the local
currency, the US dollar. In a sense, this corresponds to making a
joint analysis of the behavior of the S\&P 500 index and of the
US\$. One can worry about the possibility that something has
affected the US\$ so that the behavior of the S\&P 500 index may
have been distorted when viewed from the US\$ lens. This question
boils down in fact to the following: who are the investors moving
the market and what is the correct reference currency? In
\cite{foreign}, we have found strong evidence of fueling of the
2000 new economy bubble by foreign capital inflow. More generally,
foreigners constitute a growing part of the investment pool
\cite{albu} influencing US markets in particular with the
recycling of surpluses from Asian countries, as their moves in and
out of the market are more frequent and volatile than the major
investing US funds, due to a large current account deficit that
must be financed, fear of a weakening dollar, the impact of a
rising or decreasing dollar, and so on. Since the burst of the new
economy bubble in 2000, the Federal Reserve has decreased its
leading short term interest rate in a series of steps (see
\cite{slaving} for a detailed analysis of this Fed policy and its
relationship with the US stock market); it has been argued by many
observers that these moves may have artificially distorted the
available liquidity in addition to direct monetary interventions,
amounting to an effective inflation in dollar terms, hence its
depreciation, with observable consequences in the real estate boom
\cite{estate} (the rising price of real estate is the same as the
decrease in the value of the dollar with respect to these assets).
This suggests to deconvolve the time evolution of the US stock
market from the US\$, which amounts to taking the view point of
either a prudent investor comparing stock with a supposedly risk
haven such a gold or the view point of a foreign investor by
converting the market price in euro, British pound or Yen, for
instance.

\begin{figure}
\centering
\includegraphics[width=12cm]{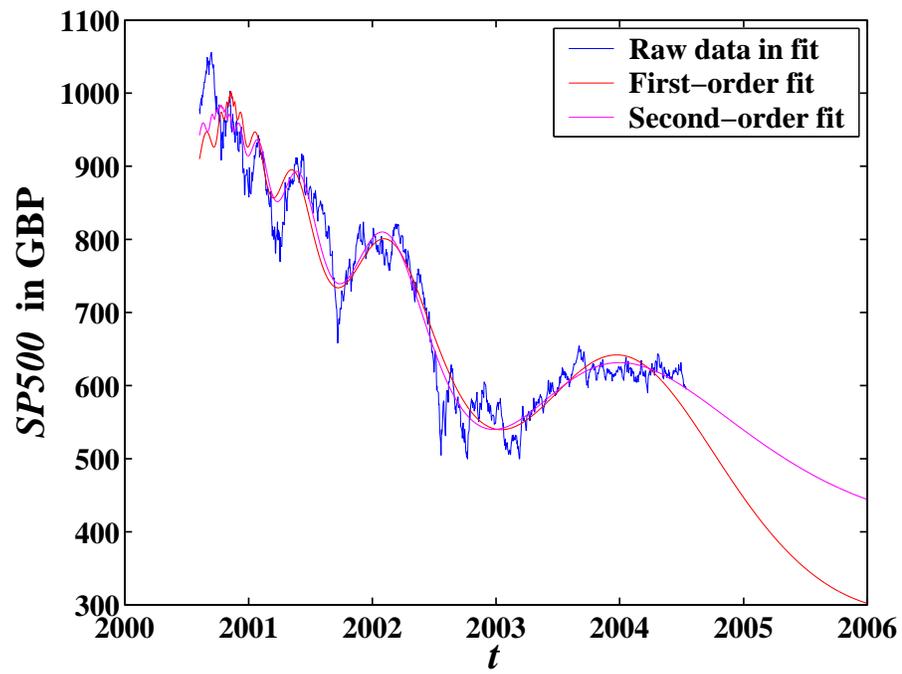}
\caption{The S\&P 500 index denominated in GBP from 2000/08/09 to
2004/07/16 and its fits using the first-order and second-order
Landau formulas. The values of the fit parameters are listed in
Table \ref{Tb3}. The fits are extrapolated to the end of 2005.}
\label{Fig:SPinGBP}
\end{figure}

We have fitted the S\&P 500 index denominated in British pound,
Canadian dollar, euro, gold fixes FM{\footnote{The price of gold
is fixed twice a day in London by the five members of the London
gold pool, all members of the London Bullion Market Association.
The fixes start at 10:30 a.m., and 3:00 p.m. London time. The data
are retrieved from http://www.amark.com/archives/fixes.asp.}},
Hong Kong dollar, Japanese Yen, XAG, as well as US dollar for
comparison, from 2000/08/09 to 2004/07/16, using the first-order
and second-order Landau formulae. We find that the fits for
Japanese Yen and XAG are even worse than that for the US dollar,
while Canadian dollar, gold fixes FM, and Hong-Kong dollar give
similar results compared with the US dollar. Interestingly, the
analyses using the British pound and the euro give much more
convincing fits. A typical plot is illustrated in
Fig.~\ref{Fig:SPinGBP} for the US market expressed in British
pounds. The parameter values of the fitting to S\&P 500
denominated in different currencies (British pound, euro, gold
fixes FM, and US dollar) are listed in Table \ref{Tb3}. Given the
quality of such fits, our previous methodology (not shown for
brevity) concludes that the antibubble continues from the European
investor view point.

\begin{figure}
\centering
\includegraphics[width=12cm]{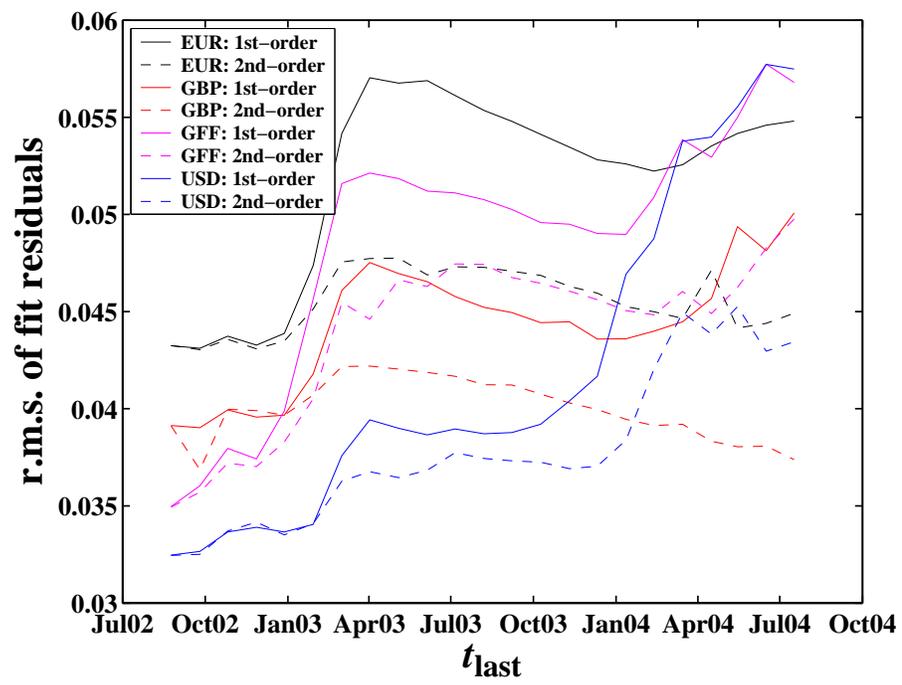}
\caption{Evolution of the RMS of residuals of the fit of the S\&P500
index expressed
in four different currencies with the
first-order and second-order Landau formulae.}
\label{Fig:RMSE}
\end{figure}

The parameters shown in Table \ref{Tb3} suggest that the crossover from
the first-order to the second-order regime has occurred, which
means a significant change in the values of the angular
log-frequency during the development of the antibubble. We note
in particular a quite significant
difference of RMSE's between the two fits, as also
shown in Fig.~\ref{Fig:RMSE}. Figure \ref{Fig:RMSE} shows the
evolution of the r.m.s. (root-mean-square, an inverse measure of
the quality of the fits) of the fit residuals of the respective
fits. In general, the discrepancy between the two fits (with the
first-order and second-order formulae) of a given
currency increases when more data are included. The separations between
the dashed versus corresponding continuous lines illustrate the
crossover from the first order to the second order.

Figure \ref{Fig:RMSE} identifies very clearly a change of
regime around February 2003, materialized by the jump in r.m.s. in
all fits and, at the same time, the sudden increase of the r.m.s. of the
first-order formula compared with the r.m.s. of the second-order formula.
The same phenomenon is documented above for the S\&P 500 in US dollar.
Beyond the quality and predictive power of the proposed fits,
we would like to stress the importance of identifying ``regime
switches''. Roughly speaking, Fig. \ref{Fig:RMSE} shows that the
r.m.s. of the fit residuals for foreign currencies of the
second-order Landau formula keep decreasing as a function of time
(the quality of the fits increase), in contrast with those of the
first-order formula. This confirms the visual impression that the
second-order Landau fits capture very well the LPPL oscillations
when compared with the first-order fits, as exemplified in
Fig.~\ref{Fig:SPinGBP}.

\begin{figure}
\centering
\includegraphics[width=12cm]{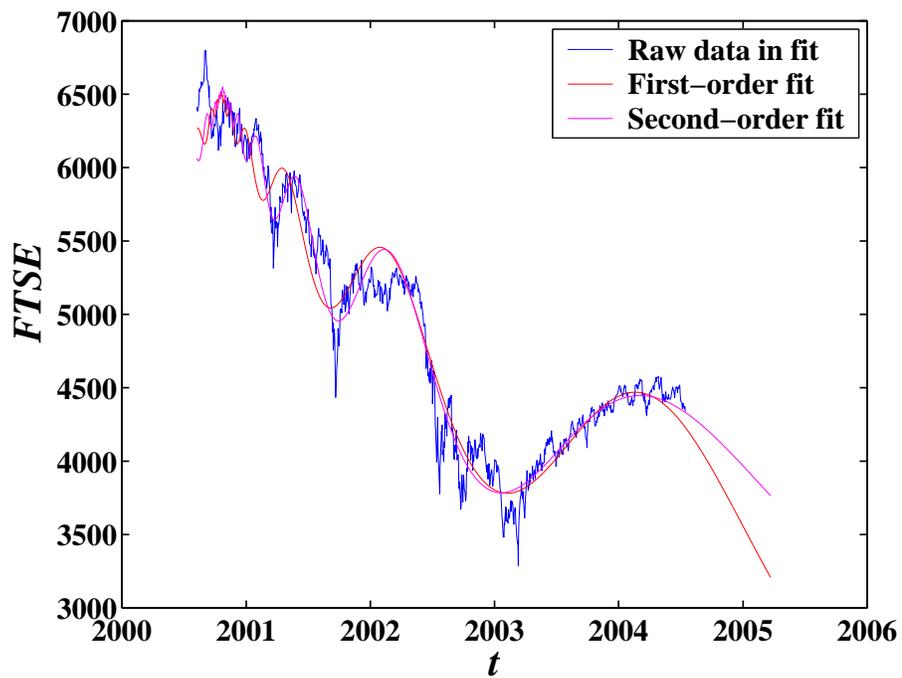}
\caption{The FTSE index (in british pounds) from 2000/08/09 to
2004/07/16 and its fits
using the first-order and second-order Landau formulas. The fits
are extrapolated to the beginning of 2005.} \label{Fig:FTSE}
\end{figure}

These analyses imply that the S\&P 500 antibubble started in
mid-2000 is still alive, when denominated in European currencies.
A natural question is then to ask if the antibubble in the
European stock markets is still continuing since those stocks are
traded directly in EUR and in GBP. We have fitted three major
indexes in Europe, that is, CAC 40 of France, DAX of Germany, and
FTSE 100 of the United Kingdom. The results are very similar to
each other. We thus take FTSE as an example shown in Figure
\ref{Fig:FTSE}. This figure presents the FTSE of the United
Kingdom from 2000/08/09 to 2004/07/16 and its fits using the
first-order and second-order Landau formulas. We see that the
antibubble is right on track in these stock markets in Europe. The
tentative conclusion of this study is that the strong impact of
the intervention of the US Federal Reserve has perturbed the
fingerprints of the antibubble of the US stock markets when viewed
in local currencies, while it is possible in reality that the
herding bearish-bullish oscillations are still present but are
hidden by the distorting feedback actions of the Federal Reserve
and the perturbed behavior of the US\$. Correcting for this
possible bias by taking the view point of an European investor, we
conclude that the antibubble may well be continuing. Similar
conclusions hold when taking gold as the reference unit to
express the value of the US stock markets (see
\cite{urlprediction}).

\section{Concluding remarks}

First, we have presented a general methodology to test for a cross-over
or a shift in log-periodicity. Second, we have developed a battery
of tests to detect a possible end of the antibubble. Our
conclusion is that the antibubble was still probably alive in
August 2003 but has ended since in the USA (i.e., when viewed from
the view point of a US investor valuing in US dollars).
More generally, our tests provide new
quantitative measures to diagnose the end of an antibubble and
this will be useful for future applications. We find that such
diagnostic is not instantaneous and requires probably three to six
months within the new regime before assessing its existence with
confidence. We have also found that the antibubble is still
continuing when viewed from the point of view of a European
investor or alternatively from an investor valuing the US stock
market with respect to gold. We attribute the discrepancy between
our two conclusions to the depreciation of the US dollar in the
last two years, which is linked to the Federal Reserve interest
rate and monetary policy.

This present paper follows several others 
\cite{SZ02QF,ZS02Global,SZ03QF,ZS03RG}
which were nucleated by noticing a similarity between
the Nikkei antibubble that started in January 1990 and the present
US antibubble that started in August 2000, when shifted approximately 
by 11 years.
We conclude with a word of caution concerning this similitude between
the two time series: the noted similarity should
not lead to the belief that the S\&P 500 index is bound to follow
blindly in the future the path suggested by the Nikkei. In
contrast with chartism or technical analysis, our approach is to
develop a scientific understanding of these bubble-antibubble
phases. The similitude between the Nikkei and US markets is part
of the search for ``universal'' properties, that allow us to
establish a theory (in short, a theory is a story of
repeatable/reproducible occurrences). Using this theory then
allows us to describe idiosyncratic behaviors, that is, deviations
from one case to another, or in other words, the parts of the
evolutions that are not universal. This is what should give us an
hedge for predictions. This is why we have emphasized in previous
works the similitude between the shifted Nikkei and the US stock
markets \cite{SZ03QF}.

However, after three-year evolution of the S\&P 500 antibubble,
the discrepancy between the Nikkei and S\&P 500 antibubbles
became detectable. The qualitative analogy is still there but,
quantitatively, there are differences. Technically, after two
years and a half after the top in December 31, 1989, we find that
the Nikkei has started to shift to another antibubble regime while
no such shift was detectable after the same time span since the
start of the antibubble in the US. Only when using data up to the
summer of 2003,
we find suggestions of such a change of regime. In addition, the
US markets have been characterized by much stronger crashes and
rallies, modelled by the so-called zero-phase Weierstrass-type
functions \cite{ZS03RG}. These two facts suggest that the herding
forces are even stronger in the US and that investors react even
more on hair-trigger to any ``news.'' The similarities between the
shifted Nikkei and the S\&P 500 are qualitative: bubble preceding
antibubble, strong speculation and herding, similar fear and
herding in the antibubble regime, some problems with bad loans or
bad accounting, strong commitment from the central banks and
governments to provide liquidity and cash... But there are
differences and these differences can be detected already after three-year
evolution of the S\&P 500 antibubble and even more after four years.

There are also interesting structural differences in the origin of
the bubbles that preceded their antibubbles. Japan was (and still
is) a surplus country, whose strong positive balance of payment
led to ``high-powered'' money being poured in the country. This in
turn powered speculation and price appreciation to sky-rocketing
levels. The so-called bad loans dragging down the Japanese
recovery came from this epoch when the high-powered money input
was used by banks to provide loans amplified by the multiplier
effect for purchases at prices often substantially larger than the
real value. In contrast, the USA has become in the last decade a
deficit country, accumulating an increasingly large negative
balance of payment with the rest of the world. The bubble that
developed in the 1990s was fuelled indirectly by the surplus
dollars accumulated by foreign countries which were re-injected in
the US in the hope of getting a reasonable return while avoiding
the risk of appreciation of their own currencies \cite{foreign}.
The bubble had also a very strong endogenous component of
self-reinforcing belief in a ``new economy,'' a characteristic
that could be matched to the faith in the Japanese miracle
underlying the Japanese bubble. Thus both the Japanese and USA
markets are strongly linked to the behavior of international
investors and central bankers and to the belief and confidence of
investors, but the specifics of the herding and over-optimism have
sometimes different origins. It remains to be seen if this will
lead to appreciable differences in the evolution of the two
antibubbles.

\textbf{Acknowledgments} We are grateful to V.F. Pisarenko for
useful remarks on the manuscript. We also thank Harald W. Ade,
Chris Couteau, Ronald Guy, and John Hunter for helpful
interactions.

\clearpage

\begin{table}
\begin{center}
\caption{\label{Tb1} Results of ex-post tests of
the end of an antibubble described in section
\ref{s2:CP} with two classes of synthetic time series.}
\medskip
\begin{tabular}{c|cccccccccccc}
\hline\hline
&$t_{\rm{last}}$&$t_c$&$m$&$\omega$&$\phi$&$A$&$B$&$C$&$\chi$\\\hline
$P_1:$&2001/08/15&0.000&0.014&0.000&0.418&0.000&0.000&0.585&0.000\\
&2002/02/15&0.001&0.000&0.000&0.053&0.000&0.000&0.000&0.000\\
&2002/08/15&0.701&0.051&0.169&0.314&0.161&0.077&0.011&0.000\\
&2003/02/15&0.139&0.945&0.011&0.081&0.065&0.969&0.894&0.000\\\hline
$P_2:$&2001/08/15&0.838&0.119&0.582&0.345&0.749&0.290&0.031&0.136\\
&2002/02/15&0.655&0.369&0.537&0.908&0.616&0.328&0.265&0.404\\
&2002/08/15&0.067&0.066&0.096&0.838&0.050&0.057&0.115&0.289\\
&2003/02/15&0.273&0.002&0.116&0.733&0.152&0.002&0.000&0.041\\\hline
$P_1^*$:&2001/08/15&0.000&0.018&0.000&0.530&0.000&0.000&0.662&0.000\\
&2002/02/15&0.000&0.000&0.000&0.347&0.000&0.033&0.000&0.000\\
&2002/08/15&0.631&0.297&0.339&0.145&0.461&0.351&0.085&0.000\\
&2003/02/15&0.236&0.982&0.030&0.080&0.070&0.993&0.943&0.001\\
\hline\hline
\end{tabular}
\end{center}
\end{table}

\clearpage

\begin{sidewaystable}
\begin{center}
\caption{\label{Tb2} Probabilities for the two types of errors
concerning seven scenarios extrapolating the S\&P 500 index over
six months from 2003/08/15 in the future. The numbers in the
parentheses stand for the standard deviations. See section
\ref{s2:ScenariosI} for a description of the testing procedure.}
\medskip
\begin{tabular}{c|cllllllll}
\hline\hline
&{\small{Scenario}}&$~~t_c$&$~~m$&$~~\omega$&$~~\phi$&$~~A$&$~~B$&$~~C$&$~~\chi$\\\hline
$P_1:$
&i&$0.000(  1)$&$0.001(  6)$&$0.000(  0)$&$0.057( 64)$&$0.000(
0)$&$0.172(216)$&$0.000(  0)$&$0.000(  0)$\\%
&ii&$0.000(  3)$&$0.000(  0)$&$0.000(  0)$&$0.093(118)$&$0.007(
35)$&$0.000(  0)$&$0.000(  0)$&$0.000(  0)$\\%
&iii&$0.025( 58)$&$0.000(  0)$&$0.002(
3)$&$0.271(242)$&$0.086(141)$&$0.000(  0)$&$0.000(  0)$&$0.004(
2)$\\%
&iv&$0.216(105)$&$0.002(  3)$&$0.057( 81)$&$0.066( 74)$&$0.058(
72)$&$0.002(  3)$&$0.019( 86)$&$0.009(  1)$\\%
&v&$0.063( 70)$&$0.000(  1)$&$0.008(
14)$&$0.100(115)$&$0.309(218)$&$0.002( 27)$&$0.000(  2)$&$0.000(
0)$\\%
&vi&$0.490(283)$&$0.508(294)$&$0.485(277)$&$0.484(280)$&$0.495(290)$&$0.508(294)$&$0.511(298)$&$0.495(284)$\\%
&vii&$0.436(261)$&$0.215(303)$&$0.378(281)$&$0.386(278)$&$0.305(250)$&$0.216(299)$&$0.213(286)$&$0.024(
82)$\\\hline%
$P_2:$
&i&$0.955( 44)$&$0.269( 37)$&$0.917( 34)$&$0.814(215)$&$0.982(
4)$&$0.061( 33)$&$0.728( 31)$&$0.965(  3)$\\%
&ii&$0.847( 67)$&$0.714( 62)$&$0.818( 57)$&$0.675(212)$&$0.843(
86)$&$0.712( 66)$&$0.898( 32)$&$0.805(  6)$\\%
&iii&$0.581(144)$&$0.665(
68)$&$0.601(101)$&$0.351(250)$&$0.467(197)$&$0.792( 80)$&$0.697(
30)$&$0.239( 26)$\\%
&iv&$0.243( 99)$&$0.184( 20)$&$0.251( 66)$&$0.701(262)$&$0.442(
90)$&$0.232( 28)$&$0.096( 17)$&$0.153( 31)$\\%
&v&$0.431(138)$&$0.420( 62)$&$0.474(
97)$&$0.726(247)$&$0.218(203)$&$0.300( 49)$&$0.374( 23)$&$0.771(
21)$\\%
&vi&$0.131(137)$&$0.026( 28)$&$0.075(
84)$&$0.191(236)$&$0.129(144)$&$0.027( 30)$&$0.025( 21)$&$0.014(
36)$\\%
&vii&$0.150(139)$&$0.084( 65)$&$0.106(
92)$&$0.258(256)$&$0.223(178)$&$0.085( 66)$&$0.065( 45)$&$0.195(
99)$\\%
\hline\hline
\end{tabular}
\end{center}
\end{sidewaystable}

\clearpage

\begin{table}
\begin{center}
\caption{\label{Tb3} Parameter values of the fitting to S\&P 500
denominated in different currencies (British pound, euro, gold
fixes FM, and US dollar) using the first-order and second-order
Landau formulae. The data are from 2000/08/09 to 2004/07/16. The
superscripts $1$ and $2$ stand for the order of the Landau formula
used in fitting.}
\medskip
\begin{tabular}{c|cccccccccccc}
\hline\hline
Currency&$t_c$&$m$&$\omega$&$\phi$&$\Delta_t$&$\Delta_\omega$&$A$&$B$&$C$&$\chi$\\\hline
EUR$^1$&2000/10/31&0.84& 6.56&3.48&     &     &7.40&-1.95E-3&5.46E-4&0.0548 \\
EUR$^2$&2000/10/05&0.90& 9.23&0.00&3343&-45&7.40&-1.29E-3&4.44E-4&0.0449 \\
GBP$^1$&2000/10/07&0.78& 7.07&0.02&     &     &6.91&-2.66E-3&7.97E-4&0.0501 \\
GBP$^2$&2000/07/30&0.99&11.92&4.05&2689&-49&6.92&-6.20E-4&-2.07E-4&0.0374 \\
GFF$^1$&2000/09/17&0.91& 5.27&2.70&     &     &1.69&-1.26E-3&-3.30E-4&0.0568 \\
GFF$^2$&2000/10/18&0.90& 8.65&0.72&5232&-78&1.69&-1.49E-3&-4.06E-4&0.0498 \\
USD$^1$&2000/09/05&0.72& 5.63&0.32&     &     &7.25&-2.55E-3&-1.20E-3&0.0575 \\
USD$^2$&2000/09/08&0.63&11.01&2.12&6902&-64&7.29&-5.87E-3&2.48E-3&0.0434\\
\hline\hline
\end{tabular}
\end{center}
\end{table}

\clearpage

\clearpage

\clearpage

\clearpage

\clearpage

\clearpage

\clearpage

\clearpage

\clearpage

\clearpage

\clearpage

\clearpage

\clearpage

\clearpage

\clearpage

\clearpage

\clearpage

\end{document}